\documentclass{openjournal}

\usepackage{lipsum}
\usepackage{amsmath} 
\usepackage{subcaption}

\usepackage[table]{xcolor}
\renewcommand{\textcolor}[2]{#2}
\usepackage{textgreek}
\usepackage[utf8]{inputenc}
\usepackage[english]{babel}

\usepackage{hyperref}
\hypersetup{
    unicode, 
    colorlinks=true,
    linkcolor=linkcolor,
    citecolor=linkcolor,
    filecolor=linkcolor,
    urlcolor=linkcolor,
}
\usepackage{color,colortbl}
\definecolor{linkcolor}{rgb}{0.0,0.3,0.5}
\usepackage{tensind}
\tensordelimiter{?}
\DeclareGraphicsExtensions{.bmp,.png,.jpg,.pdf}
\usepackage{verbatim}
\usepackage[normalem]{ulem}
\usepackage{orcidlink}
\usepackage{soul}

\graphicspath{ {./figs/} }

\makeatletter
\expandafter\let\csname longtable*\endcsname\relax
\expandafter\let\csname endlongtable*\endcsname\relax
\makeatother

\begin{document}

\title{A Population Synthesis Model for Odd Radio Circles Produced by Shock-Bubble Interactions}

\author{Yiting Wang\orcidlink{0000-0002-3143-4552}}
\email{wang2632@wisc.edu}
\affiliation{Department of Physics, University of Wisconsin-Madison}

\author{Sebastian Heinz\orcidlink{0000-0002-8433-8652}}
\email{sheinz@wisc.edu}
\affiliation{Department of Physics, University of Wisconsin-Madison}
\affiliation{Department of Astronomy, University of Wisconsin-Madison}

\author{Alexander Tchekhovskoy\orcidlink{0000-0002-9182-2047}}
\email{atchekho@northwestern.edu}
\affiliation{Department of Physics and Astronomy, Northwestern University}


\begin{abstract}
Odd Radio Circles (ORCs) are edge-brightened rings of diffuse radio emission of roughly arcminute size whose physical origin remains debated. One proposed mechanism is vortex formation through Shock-Bubble interaction (SBI), in which a merger or accretion shock re-energizes a fossil radio lobe into an edge-brightened ring. We test whether this mechanism can reproduce the observed ORC abundance using a Monte Carlo population-synthesis model that combines the galaxy stellar mass function, the radio-AGN incidence relation, dark-matter halo merger histories, and the inflation and buoyant survival of AGN-inflated bubbles, with synthetic radio fluxes and ring geometries drawn from a grid of magnetohydrodynamic SBI simulations. Applying survey-specific selection to the ASKAP-EMU Pilot Survey (944 MHz) and the LOFAR Two-metre Sky Survey (LoTSS; 144 MHz), we predict $\simeq\textcolor{red}{2.7^{+3.9}_{-1.6}}$ detectable ORCs in the EMU Pilot, matching the three found there, and $\simeq\textcolor{red}{268^{+331}_{-148}}$ in LoTSS, of which $\simeq\textcolor{red}{7.5^{+9.5}_{-4.2}}$ have a counterpart at 1.4 GHz. The predicted hosts are mostly central galaxies in group-scale halos. The predicted rings are physically smaller and lie at lower redshifts than the galaxies identified as hosts of the known ORCs, suggesting that SBI produces mainly the nearby part of the ORC population, or that some ORCs are associated with the wrong host. The population synthesis model also predicts a large population of low surface brightness steep-spectrum ORCs in the LoTSS band, implying a large hidden population and predicting that deeper low-frequency surveys should uncover many more ORCs than are currently known. These results indicate that shock-bubble interactions can account for at least part of the observed incidence of ORCs.
\end{abstract}


\maketitle

\section{Introduction}
\label{sec:intro}
Odd Radio Circles (ORCs) are a recently discovered class of faint radio sources with a circular, edge-brightened morphology and angular diameters of about one arcminute. The first ORCs were identified in observations from the Australian Square Kilometre Array Pathfinder (ASKAP) \citep{norris_unexpected_2021,norris_odd_2021}. Their large physical sizes, low brightness in radio emission, and circular morphologies are difficult to explain using standard models of radio galaxies, supernova remnants, or shocks produced directly by accretion or mergers. The small number of observed ORCs also suggests that the physical conditions and mechanisms required to produce such phenomena are uncommon. \citet{gasperin_single-_2026} searched the third data release of the LOFAR Two-metre Sky Survey (LoTSS DR3) at 144 MHz, combining catalogue filtering with visual inspection, and identified 18 sources with ORC structure: four double-headed ORCs (two of them new), five new single-headed ORCs, and nine candidates. A further candidate has been reported in the full ASKAP EMU survey \citep{filipovic_askap_2026}. Even including candidates, only a few dozen ORCs are known across these surveys, so they remain rare.

\begin{figure}
    \centering
    \begin{subfigure}[b]{0.48\columnwidth}
        \centering
        \includegraphics[height=5.5cm,keepaspectratio]{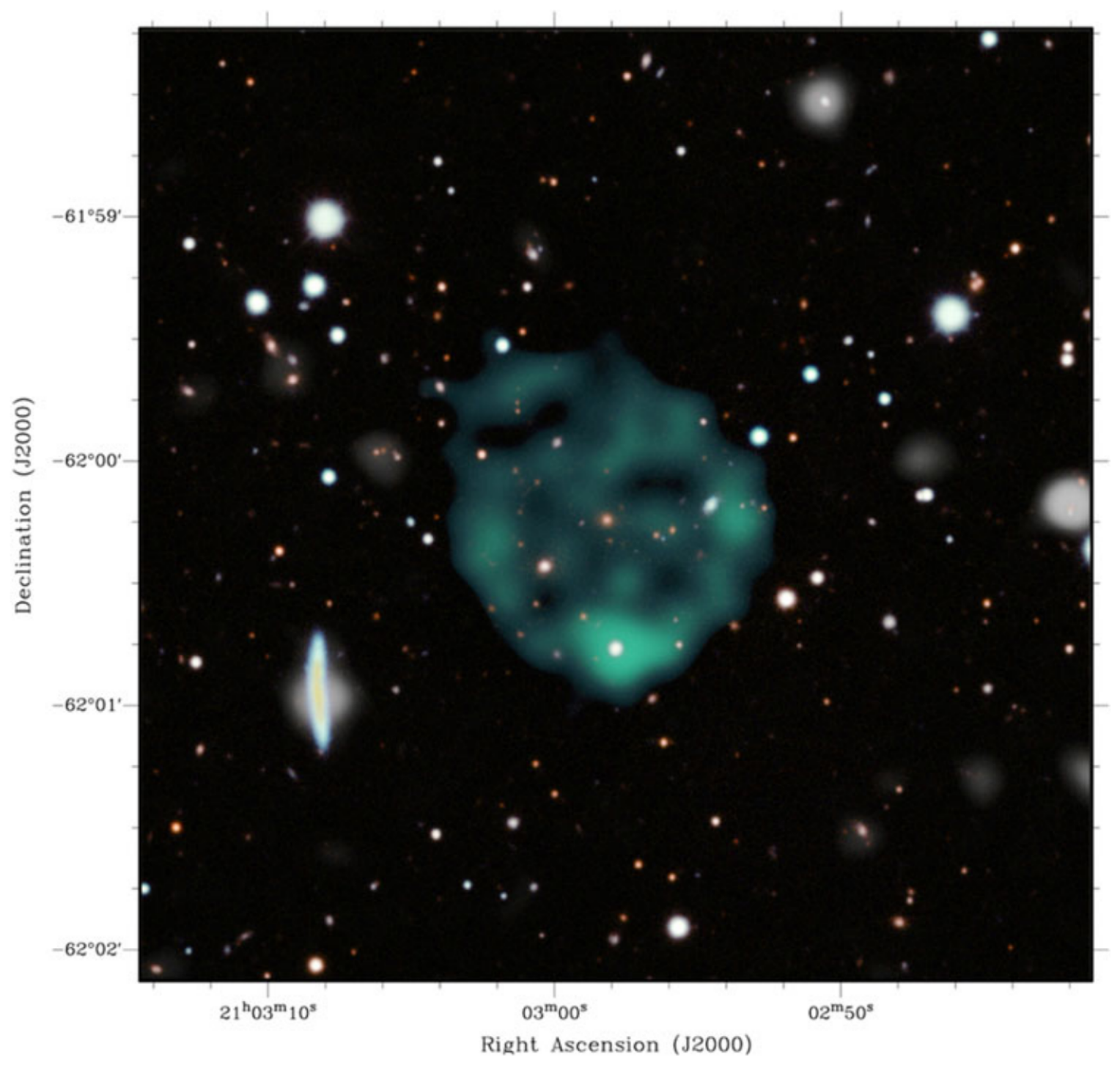}
        \caption{Observed ORC1 (ASKAP, 944 MHz)}
        \label{fig:orc1_observed_obs}
    \end{subfigure}
    \hfill
    \begin{subfigure}[b]{0.48\columnwidth}
        \centering
        \includegraphics[height=5.5cm,keepaspectratio]{SBI_simulated_ORC1.pdf}
        \caption{Simulated SBI ring}
        \label{fig:orc1_observed_sim}
    \end{subfigure}
    \caption{\textit{Left:} Observed radio emission from ORC1 at 944 MHz, overlaid on an optical background image from \citet{norris_odd_2021}. 
    \textcolor{black}{\textit{Right:} A simulated radio image of a Shock-Bubble interaction from the MHD simulations of \citet{wang_odd_2026}, showing the same edge-brightened, ring-like morphology reproduced by the model.}}
    \label{fig:orc1_observed}
\end{figure}

Progress in constraining their possible nature and physical properties relies on defining a physical model that can explain the observed structures as well as their rarity. Figure~\ref{fig:orc1_observed} (left) shows the extended, faint ring morphology of the observed ORC1. \textcolor{black}{Several formation mechanisms have been suggested. \citet{ivleva_cosmological_2026} used cosmological zoom-in simulations to show that merger-driven shocks in galaxy groups can generate ORC-like structures. \citet{lin_active_2024} investigated the possibility of ORCs as end-on views of cosmic-ray proton-dominated bubbles, and population synthesis based on their model also reproduces the observed ORC counts (Yang, priv. comm.).}

One promising explanation is that ORCs are produced when a large-scale shock interacts with a fossil radio bubble from an earlier episode of active galactic nucleus (AGN) activity \citep{shabala_are_2024}. Three-dimensional magnetohydrodynamic simulations of Shock-Bubble Interactions (SBI) have shown that this mechanism can produce the large, ring-like morphologies with spectral properties and polarization consistent with observed ORCs \citep{wang_odd_2026}. In this picture, an AGN jet first inflates a low-density, magnetized bubble filled with non-thermal plasma. After the AGN episode, the bubble stays and evolves passively until it encounters a (likely merger-driven\footnote{Other origins for shocks in this process include shocks driven by subsequent AGN activity.}) shock. The shock compresses the bubble into a vortex ring and re-energizes the relativistic electrons through first-order Fermi acceleration, adiabatic compression, and turbulent re-acceleration, producing synchrotron emission at observable frequencies. When viewed from a favorable direction, the compressed bubble can appear as an edge-brightened radio ring in the low-pressure group environments.

The simulations have established that SBI is a physically plausible mechanism for producing individual ORCs, however, they do not explain whether such events occur frequently enough to explain the observed population. The model requires several conditions combined. A galaxy must undergo an AGN jet episode capable of inflating large bubbles of non-thermal plasma (as observed in classical radio galaxies). The number and powers of AGN jets depend on the galaxy mass and the radio-AGN incidence \citep{best_host_2005,sabater_lotss_2019,igo_lofar_2024,oshea_shooting_2025}. The bubble energy and size depend on the jet power, jet lifetime, and ambient gas pressure \citep{birzan_systematic_2004,osullivan_heating_2011,hardcastle_radio-loud_2019}. The bubble must then 
remain intact until a (merger) shock reaches its location. The environmental pressure depends on the galaxy's host halo mass and position within the halo \citep{yang_galaxy_2009,tinker_toward_2008,budzynski_radial_2012}. The probability of encountering a merger shock depends on the halo merger history \citep{fakhouri_merger_2010}. The host halo must undergo a merger to generate a shock of the appropriate strength. Merger shocks themselves span distributions of ages, surface areas, locations, and Mach numbers \citep{vazza_shock_2009,ha_properties_2018,lee_morphology_2024}. Finally, the resulting vortex ring must have the angular size and surface brightness that are large enough to be detectable as an ORC by a radio survey. 

In this work, we construct end-to-end Monte Carlo population synthesis simulations to predict the incidence of ORCs produced through SBI. This paper attempts to explain the observable population and tests whether the SBI can be consistent with the small number of ORCs currently known. 

This paper is organized as follows. In Section~\ref{sec:methods}, we describe the entire population synthesis framework, starting with the galaxy and host-halo assignment, going through shock and merger histories and the AGN episodes that inflate the bubbles, ending with applying observational selections that determines which bubbles are detectable as ORCs. In Section~\ref{sec:results}, we present the resulting distributions of host galaxies, halo environments, shock merger distributions, AGN episodes with fossil bubbles, and the predicted intrinsic and observable ORC incidence. In Section~\ref{sec:discussion}, we compare the model's predictions with the observed population and discuss the uncertainties and limitations of the calculation. We summarize our conclusions in Section~\ref{sec:concl}.

\section{Population Synthesis Model}
\label{sec:methods}
Our goal is to predict how many ORCs an ASKAP/EMU-like survey would detect if they were produced by merger shocks encountering fossil AGN lobes (referred to as bubbles below). Because the outcome depends on a long chain of conditions, we build a Monte Carlo method to construct the incidence function. 
We describe the chain briefly here, while we provide the full technical details of each modeling step in Sections~\ref{sec:stellar_mass_radioagn}--\ref{sec:flux_grid}. Every step draws from an empirically calibrated distribution and carries a comoving number density weight, so that the final count can be converted into an expected number within the survey volume.
The calculation steps follow the sequence below:
\begin{enumerate}
  \item \textbf{Galaxies and their radio AGN} (Section~\ref{sec:stellar_mass_radioagn}). 
We draw $N_{\rm gal}$ stellar masses from the local stellar mass function of \citet{bernardi_stellar_2018}. We neglect cosmic evolution of the galaxy mass function, given the relatively low redshifts inferred for the observed sample of ORCs, with $z\lesssim 0.55$. For a given galaxy, we then decide whether the galaxy hosts a radio AGN and what its jet power is, using the incidence relation measured by \citet{igo_lofar_2024} and parameterized by \citet{oshea_shooting_2025}.

  \item \textbf{AGN host halos} (Section~\ref{sec:halo}). 
Each selected AGN is assigned a dark matter halo by combining the conditional stellar mass function of \citet{yang_galaxy_2009} with the halo mass function of \citet{tinker_toward_2008}.

  \item \textbf{Merger histories} (Section~\ref{sec:merger}). 
We use the conditional merger rate measured by \citet{fakhouri_merger_2010}. We draw the two most recent major mergers of each halo. \textcolor{red}{We assume that only the most recent merger shock can hit the bubbles and produce an ORC that is observable today. The second most recent shock sets the earliest time at which such a bubble can be inflated. Any bubble inflated before it was already hit by that shock and do not count toward the observation today. We assume that when the most recent shock later passes through this old ring, it does not form a new one, because the bubble is no longer a coherent underdense cavity.} Thus, the interval between the two most recent shocks is the window during which we consider inflation of progenitor bubbles. 

  \item \textbf{Progenitor environment} (Sections~\ref{sec:progmass}--\ref{sec:progenv}).
The radio-AGN host galaxy selected in step~1 is placed into the progenitor halo. The conditional stellar mass function of \citet{yang_galaxy_2009} is used to classify it as a central or satellite galaxy. Satellites are placed at a halo-centric radius following \citet{budzynski_radial_2012}. A $\beta$-model atmosphere at the halo's virial temperature, normalized to the gas fraction of \citet{popesso_hot_2026}, then gives the ambient pressure at any position in the halo, which sets how large a bubble of given energy can grow.

  \item \textbf{AGN episodes and bubbles} (Sections~\ref{sec:episodes}).
We consider a simple, heuristic model for bubble inflation. During the interval between the two mergers, we follow the AGN activity to generate a chain of AGN outbursts, corresponding to the AGN undergoing changes in mean jet power. Consecutive episodes energize the same bubble/cavity until the cavity leaves the host galaxy, after which a new bubble is generated. \textcolor{red}{The host separates from its bubble through its own motion in the halo and the buoyant rise of the bubble. The bubble rises away from the halo center rather than following the host. We combine the two into a single separation speed. The bubble is closed once the host has moved farther than the bubble radius, and the next episode starts a new one.} Each bubble is inflated next to its host (mimicking jet-inflated cocoons), and its size is solved for with the gas pressure at its own position. After the jet switches off, the bubble rises \textcolor{red}{buoyantly away from the galaxy and continues to expand. Bubbles that rise more than three times their radius before the shock arrives are considered disrupted and removed (a fiducial choice from the simulations of \citet{churazov_evolution_2001}); the others are hit by the shock.}

  \item \textbf{Shock strength} (Section~\ref{sec:mach}).
The most recent merger shock is assigned a Mach number drawn from the distribution measured in cluster-merger simulations by \citet{lee_morphology_2024}. If the Mach number lies within the range suitable for vortex generation \citep{wang_odd_2026}, the bubbles inflated during the window bounded by the two most recent mergers are considered for ORC generation.

  \item \textbf{Observability} (Section~\ref{sec:survey}).
The radio flux of every vortex ring created from a shocked bubble is computed by rescaling a pre-computed grid of MHD shock-bubble simulations \citep{wang_odd_2026}, and angular-size, integrated-flux and surface-brightness cuts appropriate to the EMU Pilot survey and the Lofar LOTSS are applied. Lastly, a viewing-angle factor converts the number of detectable bubbles into the expected number of ORC-like sources.
\end{enumerate}

Throughout, masses are quoted in physical units ($M_\odot$), lengths in kpc unless stated otherwise, and all number densities are comoving.

\subsection{Stellar mass and radio AGN sampling}
\label{sec:stellar_mass_radioagn}
The first step is to construct a sample of galaxies and determine which of them host a radio AGN with a given jet power. 
We draw galaxy stellar masses from the local galaxy stellar mass function measured by \citet{bernardi_stellar_2018}. We use their Table~B2 values for SerExp photometry under a fixed Chabrier IMF (the \texttt{M14d}/\texttt{M14df} columns, dusty and dust-free respectively). We adopt the midpoint of these two columns as our mass function. A fixed IMF is used so that our stellar masses share the same universal-IMF convention as the \citet{yang_galaxy_2009} that we will use later for halo assignment in Section \ref{sec:halo_merger_histories}.

The stellar mass function is expressed as 
\begin{equation}
\Phi_\star(\log M_\star) \equiv \frac{dn_{\rm gal}}{d\log_{10}M_\star}
\end{equation}

Table~B2 is tabulated in $0.1$ dex bins. We interpolate $\log_{10}\Phi_\star$ linearly onto a grid of 5000 points spanning the full range of the table, $9.05 \leq \log_{10}M_\star \leq 12.25$, and evaluate
\begin{equation}
\Phi_\star(\log M_\star) = 10^{ \log_{10}\Phi_\star}
\end{equation}
onto that grid. We interpolate only within the tabulated range and never extrapolate beyond it, so the stellar masses we draw are always within this range.

The total comoving number density is
\begin{equation}
n_{\rm gal} = \int_{\log M_{\star,\min}}^{\log M_{\star,\max}} \Phi_\star(\log M_\star)  d\log_{10}M_\star \simeq 1.30\times10^{-2} {\rm Mpc^{-3}}
\end{equation}

We convert the stellar mass function into a normalized probability
density,
\begin{equation}
p(\log M_\star) = \frac{\Phi_\star(\log M_\star)}{n_{\rm gal}}
\end{equation}
and draw $N_{\rm gal}=10^6$ galaxy stellar masses by inverse transform sampling following the standard Monte Carlo sampling approach used in merger-tree calculations \citep{lacey_merger_1993,somerville_how_1999,parkinson_generating_2007}. If
\begin{equation}
F_M(\log M_\star) = \int_{\log M_{\star,\min}}^{\log M_\star} p(\log M_\star') d\log M_\star'
\end{equation}
is the cumulative distribution, a uniform random variate $u\in[0,1]$ is mapped to a stellar mass through $\log M_\star=F_M^{-1}(u)$. We evaluate $F_M$ on the same 5000-point grid and rescale it by its final value so that it runs exactly from 0 to 1.

Each Monte Carlo galaxy represents the same comoving number-density weight,
\begin{equation}
w_{\rm gal} = \frac{n_{\rm gal}}{N_{\rm gal}}
\label{eq:wgal}
\end{equation}
For $N_{\rm gal}=10^6$, this gives $w_{\rm gal} \simeq 1.30\times10^{-8}\ {\rm Mpc^{-3}}$ per sampled galaxy.

For each sampled galaxy, we next determine whether it hosts a radio AGN and, if active, draw its kinetic Eddington fraction $\lambda_{\rm jet}$ (the ratio of jet power to Eddington power.) To this end, we use the radio-AGN incidence measured from the LOFAR-eFEDS field by \citet{igo_lofar_2024} and parameterized by \citet{oshea_shooting_2025}. From the Eddington fraction we calculate the total jet power from the relation between stellar mass and black hole mass assumed by \citet{igo_global_2025}

In their eq.(9), \citet{oshea_shooting_2025} approximate the combined (compact and complex morphology) radio-AGN incidence by a plane in $\log M_\star$ and $\log\lambda_{\rm jet}$:
\begin{equation}
\label{eq:oshea}
\log_{10}\eta = a_\eta \log_{10}\left(\frac{M_\star}{M_\odot}\right) + b_\eta\log_{10}\lambda_{\rm jet} + c_\eta
\end{equation}
with
\begin{equation}
a_\eta=1.55,\qquad b_\eta=-0.845,\qquad c_\eta=-21.7
\end{equation}
Equivalently,
\begin{equation}
\eta(\log M_\star,\log\lambda_{\rm jet}) = 10^{ a_\eta\log M_\star+ b_\eta\log\lambda_{\rm jet}+ c_\eta}
\end{equation}
Here, $\eta$ has units of ${\rm dex^{-1}}$ in $\log_{10}\lambda_{\rm jet}$. Thus, $\eta d\log_{10}\lambda_{\rm jet}$ is the probability that a galaxy with stellar mass $M_\star$ hosts a radio AGN in the interval $d\log_{10}\lambda_{\rm jet}$. 

For a galaxy of stellar mass $M_\star$, the probability of hosting a radio AGN anywhere over the interval $-6 \leq \log_{10}\lambda_{\rm jet} \leq -1$ is
\begin{align}
P_{\rm AGN}(M_\star) & = \int_{\log\lambda_{\min}}^{\log\lambda_{\max}} \eta(\log M_\star,\log\lambda_{\rm jet})  d\log_{10}\lambda_{\rm jet}\\
& = 10^{a_\eta\log M_\star+c_\eta}\left(10^{b_\eta\log\lambda_{\max}}-10^{b_\eta\log\lambda_{\min}}\right)/\left(b_\eta\ln 10\right)
\label{eq:P_AGN}
\end{align}

The fitted plane continues to rise with stellar mass and formally exceeds unity above $\log_{10}M_\star\approx10.92$, so we cap the incidence at
\begin{equation}
P_{\rm AGN} = \min\left[P_{\rm AGN}(M_\star), 1\right]
\label{eq:pagn_cap}
\end{equation}
Above this mass every sampled galaxy is therefore selected as a radio AGN. We return to this extrapolation in Section~\ref{sec:caveats}.

A galaxy is selected as a radio AGN if a uniform random variate $u_{\rm sel}\in[0,1]$ satisfies
\begin{equation}
u_{\rm sel} < P_{\rm AGN}(M_\star)
\end{equation}

For galaxies selected as radio AGN, we draw $\log\lambda_{\rm jet}$ from the conditional distribution
\begin{equation}
p(\log\lambda_{\rm jet}\mid M_\star,{\rm AGN}) = \frac{ \eta(\log M_\star,\log\lambda_{\rm jet}) }{P_{\rm AGN}(M_\star)}
\end{equation}

We sample this distribution analytically by inverse transform. Let $x=\log_{10}\lambda_{\rm jet}$, with bounds $x_{\min}$ and $x_{\max}$. Its cumulative distribution is
\begin{equation}
F_\lambda(x) = \frac{ 10^{b_\eta x} - 10^{b_\eta x_{\min}} }{ 10^{b_\eta x_{\max}} - 10^{b_\eta x_{\min}} }
\end{equation}

Setting $F_\lambda(x)=u$, where $u$ is uniform on $[0,1]$, gives
\begin{equation}
\log_{10}\lambda_{\rm jet} = \frac{1}{b_\eta} \log_{10} \left[ 10^{b_\eta x_{\min}} + u \left(10^{b_\eta x_{\max}} - 10^{b_\eta x_{\min}} \right) \right]
\end{equation}

Each sampled galaxy carries a weight $w_{\rm gal}=n_{\rm gal}/N_{\rm sample}$, where $n_{\rm gal}$ is the comoving number density of galaxies in the sampled stellar-mass range at $z=0$. This weight converts the finite Monte Carlo sample into a physical number density. The comoving number density of the selected radio AGN is then the sum of their weights,
\begin{equation}
n_{\rm radioAGN} = \sum_{i\in{\rm AGN}}w_{{\rm gal},i} = N_{\rm AGN}w_{\rm gal}
\end{equation}
We assume that this comoving density does not evolve over the survey range, $0.01<z<0.55$. Each descendant halo is then placed at an observation redshift $z_0$ drawn uniformly in comoving volume, and its merger history is traced back from $z_0$. The expected number of sources in a survey is therefore $N=n V_{\rm survey}$, with no further integral over an evolving source density. 
For the fiducial $10^6$-galaxy realization, approximately $1.41\times10^5$ galaxies are selected as radio AGN, corresponding to
\begin{equation}
n_{\rm radioAGN} \simeq 1.83\times10^{-3}\ {\rm Mpc^{-3}}
\end{equation}

\textcolor{black}{This is showing that $\sim\!14\%$ of the sampled galaxies are radio-AGNs. The fraction is defined over the sampled mass range $9.05\le\log_{10}(M_\star/M_\odot)\le12.25$. The fraction depends on where the lower mass limit is placed, rising to $\simeq34\%$ for a cut at $\log_{10}M_\star=10$ and falling if the limit is lowered further. The sampled range is set by the validity of the adopted incidence relation \citet{igo_global_2025} and SMF \citet{bernardi_stellar_2018}, and the galaxies selected as radio-AGN hosts are in any case dominated by massive hosts (median $\log_{10}M_\star\simeq10.8$). }

The incidence relation of \citet{igo_lofar_2024} is measured from a spectroscopic sample at $z<0.4$ and is applied here without redshift evolution, as are the stellar mass function and the halo statistics of Section~\ref{sec:halo_merger_histories}. We discuss this assumption in Section~\ref{sec:caveats}.

\begin{figure*}[t]
    \centering
    \includegraphics[width=0.5\columnwidth]{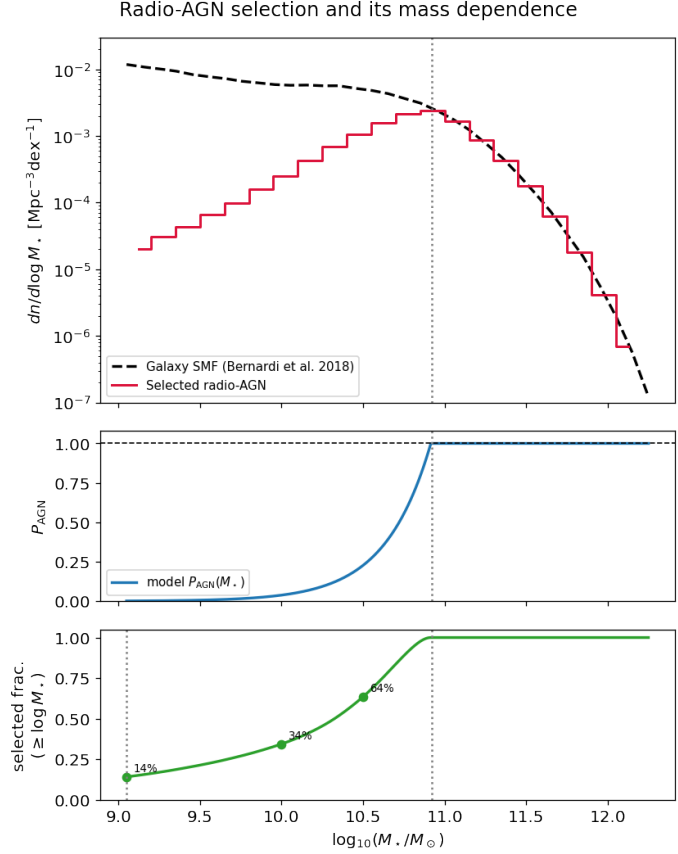}
    \caption{
    \emph{Top:} The galaxy stellar mass function (dashed black) adopted from Table~B2 of \citet{bernardi_stellar_2018}; the Monte Carlo sampled $10^6$ galaxies (solid gray) overlapping the black line; and the resulting selected radio-AGN population (red). \emph{Middle:} The incidence probability $P_{\rm AGN}(M_\star)$ used to select the radio-AGN. Probability reaches $1$ at $\log M_\star\approx10.92$ (vertical dotted line, shown in both panels). \emph{Bottom:} The cumulative selected fraction starting from the lower stellar-mass limit, rising from $\simeq14\%$ for the full sampled range to $\simeq34\%$ and $\simeq64\%$ for cuts at $\log_{10}M_\star=10$ and $10.5$.}
    \label{fig:smf_incidence_agnhosts}
\end{figure*}

Figure~\ref{fig:smf_incidence_agnhosts} shows the AGN selection process in Section~\ref{sec:stellar_mass_radioagn}. The selected AGN (in red) trace the stellar mass function closely at high masses, where $P_{\rm AGN}$ saturates at unity, and essentially every sampled galaxy (in gray) is selected. It falls below the stellar mass function toward lower masses where where $P_{\rm AGN}$ drops and only a fraction of sampled galaxies qualify as radio AGN. The transition happens at $\log M_\star\approx10.92$. Overall, 14\% of the sampled galaxies are selected as radio AGN, rising to 34 and 64\% above $\log_{10}M_\star=10$ and 10.5. Note, we use ``radio AGN'' in a broad sense. Any galaxy with a kinetic jet Eddington ratio $\lambda_{\rm jet}\geq10^{-6}$, the lower bound of the incidence relation of \citet{igo_global_2025} is considered a radio AGN here, which includes many AGN that would not be considered radio loud by the classical definition based on flux density. Most of these systems have weak jets and are not radio-loud AGN in the classical sense, but their jets still inflate bubbles that can later form ORCs.

\subsection{Stellar to host halo assignment}
\label{sec:stellar_to_halo}\label{sec:halo}
The next step is to assign a dark-matter host halo to each radio AGN selected in Section~\ref{sec:stellar_mass_radioagn}. Galaxies with the same $M_\star$ may be central galaxies in halos or satellites embedded in more massive groups and clusters. 

We construct the conditional probability distribution
\begin{equation}
p(M_{\rm h}\mid M_\star)
\end{equation}
using the conditional stellar mass functions measured by \citet{yang_galaxy_2009} together with the halo mass function of \citet{tinker_toward_2008}. We then draw one host halo mass for every selected radio AGN.

\citet{yang_galaxy_2009} constructed a galaxy-group catalog from the Sloan Digital Sky Survey and measured the distribution of galaxy stellar masses within halos of different masses. Their conditional stellar mass function (CSMF) is 
\begin{equation}
\Phi(M_\star\mid M_{\rm h})
\equiv
\frac{d\langle N_{\rm gal}\rangle}{d\log_{10}M_\star}
\end{equation}
where $\Phi(M_\star\mid M_{\rm h}) d\log_{10}M_\star$ is the mean number of galaxies in a halo of mass $M_{\rm h}$. They separate the CSMF into central- and satellite-galaxy contributions from their Eq.(13), (14), and (16)
\begin{equation}
\Phi(M_\star\mid M_{\rm h})
=
\Phi_{\rm cen}(M_\star\mid M_{\rm h})
+
\Phi_{\rm sat}(M_\star\mid M_{\rm h})
\end{equation}
which we use together as a single combined CSMF throughout this step.
The central term is a log-normal in stellar mass,
\begin{equation}
\Phi_{\rm cen}(M_\star\mid M_{\rm h})
=
\frac{1}{\sqrt{2\pi} \sigma_c}
\exp\left[-\frac{\left(\log_{10}M_\star-\log_{10}M_{\star,c}\right)^2}{2\sigma_c^2}\right]
\label{eq:csmf_cen}
\end{equation}
and the satellite term is a modified Schechter function,
\begin{equation}
\Phi_{\rm sat}(M_\star\mid M_{\rm h})
=
\ln(10) \phi_s^\star 
x^{ \alpha_s+1}\exp\left(-x^2\right),
\qquad
x \equiv \frac{M_\star}{M_{\star,s}}
\label{eq:csmf_sat}
\end{equation}
where the satellite characteristic mass is tied to the central one by
\begin{equation}
\log_{10}M_{\star,s} = \log_{10}M_{\star,c} - 0.25
\end{equation}

The four parameters $\phi_s^\star$, $\alpha_s$, $\log_{10}M_{\star,c}$, and $\sigma_c$ are tabulated in nine halo-mass bins spanning $12.16\leq\log_{10}(M_{\rm h}/h^{-1}M_\odot)\leq14.58$ in Table~4 of \citet{yang_galaxy_2009}. We interpolate them linearly in $\log_{10}M_{\rm h}$ between the tabulated bins, interpolating $\phi_s^\star$ in logarithmic space since it spans two orders of magnitude across the table.

The CSMF tells the likelihood of finding a galaxy of a particular stellar mass in a halo of fixed mass $M_{\rm h}$. The different masses of halo distribution is given by the halo mass function \citep{tinker_toward_2008} 
\begin{equation}
\frac{dn_{\rm h}}{dM_{\rm h}}
=
f(\sigma) \frac{\rho_{\rm m}}{M_{\rm h}} \frac{d\ln\sigma^{-1}}{dM_{\rm h}}
\end{equation}
where $\rho_{\rm m}$ is the mean matter density, $\sigma(M_{\rm h})$ is the linear rms density fluctuation on the mass scale $M_{\rm h}$, $f(\sigma)$ is the multiplicity function
\begin{equation}
f(\sigma) = A \left[ \left( \frac{\sigma}{b} \right)^{-a} +1 \right] \exp \left( -\frac{c}{\sigma^2} \right)
\end{equation}
with parameters $A$, $a$, $b$, and $c$ at the $\Delta=200{\rm m}$ overdensity we adopt throughout the analysis.

We evaluate the Tinker mass function at $z=0$ using the $M_{200{\rm m}}$ definition, where
\begin{equation}
M_{200{\rm m}} = \frac{4\pi}{3} R_{200{\rm m}}^3 \left(200\rho_{\rm m}\right)
\end{equation}
Thus, $R_{200{\rm m}}$ encloses a mean density equal to 200 times the cosmic mean matter density over the range of Yang's table, using the \textsc{Colossus} cosmology package \citet{diemer_colossus_2018} 
\begin{equation}
h=0.73,\qquad \Omega_{\rm m}=0.238,\qquad \Omega_{\rm b}=0.0416,\qquad \sigma_8=0.75,\qquad n_{\rm s}=0.951
\end{equation}.

The Tinker function returns $dn_{\rm h}/d\ln M_{\rm h}$. We then convert this to a number density per dex 
\begin{equation}
\frac{dn_{\rm h}}{d\log_{10}M_{\rm h}}
=
\ln(10) \frac{dn_{\rm h}}{d\ln M_{\rm h}}
\end{equation}
We treat the assigned halo mass as $M_{200{\rm m}}$ entering Yang's CSMF.

Table~4 of \citet{yang_galaxy_2009} is tabulated in $h^{-1}M_\odot$, while we work in physical units throughout. Halo masses carry one power of $h$ and stellar masses two, so
\begin{equation}
\log_{10}M_{\rm h} = \log_{10}\left(M_{\rm h}/h^{-1}M_\odot\right) - \log_{10}h,
\qquad
\log_{10}\left(M_\star/h^{-2}M_\odot\right) = \log_{10}M_\star + 2\log_{10}h
\end{equation}
We evaluate the CSMF and the halo mass function on a grid of 1000 points in $\log_{10}M_{\rm h}$ spanning the full range of Table~4, and do not extrapolate beyond it.

For a galaxy of stellar mass $M_\star$, the probability that it lives in a halo of mass $M_{\rm h}$ ($M_{\rm h}\geq M_\star$) is
\begin{equation}
p(\log M_{\rm h}\mid\log M_\star) = \frac{ \displaystyle \frac{dn_{\rm h}}{d\log M_{\rm h}} \Phi(M_\star\mid M_{\rm h}) }{ \displaystyle \int \frac{dn_{\rm h}}{d\log M_{\rm h}'} \Phi(M_\star\mid M_{\rm h}')  d\log M_{\rm h}' }
\label{eq:phalo}
\end{equation}

For each selected radio AGN, we evaluate $p(\log M_{\rm h}\mid\log M_\star)$ on the halo-mass grid, build its cumulative distribution,
\begin{equation}
F_{\rm h}(\log M_{\rm h}\mid M_\star)
= \int_{\log M_{{\rm h},\min}}^{\log M_{\rm h}} p(\log M_{\rm h}'\mid M_\star)  d\log M_{\rm h}'
\end{equation}
draw a uniform variate \(u_{\rm h}\in[0,1]\), and assign
\begin{equation}
\log M_{\rm h}
=
F_{\rm h}^{-1}(u_{\rm h}\mid M_\star)
\end{equation}
This procedure is repeated independently for each of the $141{,}167$ selected radio AGN, giving a median assigned halo mass of $\log_{10}(M_{\rm h}/M_\odot)\simeq12.2$ in the catalog.

Both the CSMF and the halo mass function are evaluated at $z=0$, while the halo structure in Section~\ref{sec:progenitor_environment} is evaluated at the epoch of the shock. Equation~\eqref{eq:phalo} uses the product of the two only as a normalized shape in $M_{\rm h}$. We return to this in Section~\ref{sec:caveats}.

\begin{figure*}[t]
    \centering
    \includegraphics[width=0.5\textwidth]{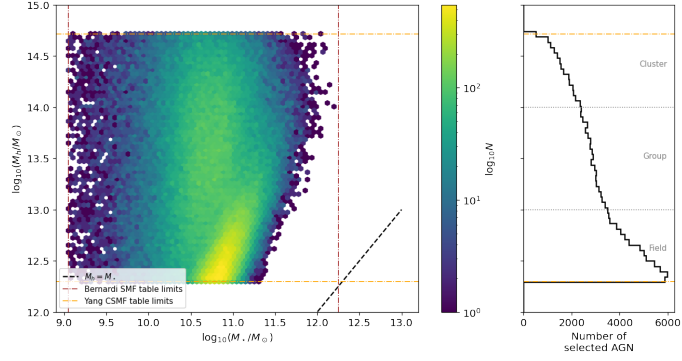}
    \caption{Assigned halo mass $M_{\rm h}$ versus stellar mass $M_\star$ for the selected radio-AGN sample. Left: Joint distribution, colored by the number of samples $\log_{10}N$ of $\log_{10}M_\star$ and the assigned $\log_{10}M_{\rm h}$. The black dashed line is the condition $M_h=M_\star$. The red dashed-dotted lines show the stellar mass function boundaries from \citet{bernardi_stellar_2018}; the orange dashed-dotted lines show the conditional stellar mass function boundaries from \citet{yang_galaxy_2009}. Right: The distribution of assigned halo masses, with the three environment categories labeled.}
    \label{fig:halomass_joint}
\end{figure*}
Figure~\ref{fig:halomass_joint} shows the results of the halo-mass assignment described above, with the boundaries of the data we used. We only consider the cases $M_h>M_\star$. The vertical red lines mark the full range of the \citet{bernardi_stellar_2018} stellar mass function used to draw galaxy masses in the first place ($9.05\leq\log_{10}M_\star\leq12.25$). The horizontal orange lines mark the actual range spanned by \citet{yang_galaxy_2009}'s Table 4, $12.16\leq\log_{10}(M_h/h^{-1}M_\odot)\leq14.58$, converted to physical units. The right-hand panel shows a histogram of the number of selected AGNs with their host halo masses. The halo mass thresholds for environment (field, group, cluster) from \citet{mccarty_strong_2025} ($\log_{10}M_h=12.0, 13.0, 14.0$) are labeled.

\subsection{Halo merger histories}
\label{sec:halo_merger_histories}\label{sec:merger}
Using the empirical distribution of halo masses we constructed in Section \ref{sec:stellar_to_halo}, we draw a random halo mass $M_{\rm h}(z) \equiv M_0$ and assign this as the descendant AGN host halo. Each descendant halo is assigned its own observation redshift $z_0$, drawn uniform in comoving volume over the survey range ($0.01<z_0<0.55$, $270\ {\rm deg}^2$; Section~\ref{sec:survey}). 
The cumulative distribution used to draw $z_0$ is
\begin{equation}
F_z(z_0) = \frac{V_{\rm c}(z_0)-V_{\rm c}(z_{\min})}{V_{\rm c}(z_{\max})-V_{\rm c}(z_{\min})}
\label{eq:zdraw}
\end{equation}
where $V_{\rm c}$ is the comoving volume, and $z_0=F_z^{-1}(u_z)$ for a uniform variate $u_z\in[0,1]$. All lookback times below are measured from this halo's own $z_0$.
We use the mean merger rate per descendant halo measured by \citet{fakhouri_merger_2010}, the number of mergers per halo per unit progenitor mass ratio and per unit redshift,
\begin{equation}
\frac{dN_{\rm merger}}{d\xi dz}
=
A \left( \frac{M_0}{10^{12}M_\odot} \right)^\alpha \xi^\beta \exp \left[ \left( \frac{\xi}{\widetilde{\xi}} \right)^\gamma \right] (1+z)^\eta
\label{eq:mergerrate}
\end{equation}
with $(A,\alpha,\beta,\gamma,\eta,\widetilde\xi)=(0.0104, 0.133, -1.995, 0.263, 0.0993, 9.72\times10^{-3})$. Here $M_0$ is the descendant halo mass at the merger event and $\xi\equiv M_{\rm small}/M_{\rm large}$ is the progenitor mass ratio ($0<\xi\leq1$), with $q\equiv1/\xi$. We restrict the calculation to major mergers, $0.25\leq\xi\leq1$ (progenitor mass ratios between $1:4$ and $1:1$), the same range analyzed in the merger-shock simulations of \citet{lee_morphology_2024}. We hold $M_0$ fixed while integrating over the halo's own history, neglecting its growth; the mass dependence in Equation~\eqref{eq:mergerrate} is weak ($\alpha=0.133$), and we return to this in Section~\ref{sec:caveats}.

The normalized probability density for the mass ratio of a major merger is
\begin{equation}
p(\xi)
=
\frac{\xi^\beta \exp \left[ \left( \xi/\widetilde{\xi} \right)^\gamma \right]}{I_\xi},
\qquad
I_\xi
=
\int_{\xi_{\min}}^{\xi_{\max}} \xi^\beta \exp \left[ \left( \xi/\widetilde{\xi} \right)^\gamma \right] d\xi
\label{eq:pxi}
\end{equation}
evaluated numerically on a grid of 5000 values in $\xi$. Its cumulative distribution 
\begin{equation}
F_\xi(\xi)
=
\int_{\xi_{\min}}^\xi p(\xi') d\xi'
\end{equation}
$F_\xi(\xi)$ is inverted at a uniform random number $u_\xi\in[0,1]$ to draw a mass ratio, $\xi=F_\xi^{-1}(u_\xi)$.

Integrating the merger rate over the adopted major-merger interval gives the mean number of major mergers per halo per unit redshift,
\begin{equation}
\frac{dN_{\rm major}}{dz}
=
A \left(\frac{M}{10^{12}M_\odot}\right)^\alpha (1+z)^\eta I_\xi
\end{equation}
The standard dark matter halo merger tree uses the Press-Schechter formalism and its extended Press-Schechter (EPS) generalization \citep{press_formation_1974,bond_excursion_1991}. In EPS models, conditional progenitor mass functions and transition probabilities can be used to construct stochastic halo assembly histories \citep{lacey_merger_1993}. We use the same Monte Carlo merger-tree algorithms, which sample progenitor masses and merger events from the corresponding conditional distributions \citep{somerville_how_1999,parkinson_generating_2007}, but use the simulation-calibrated merger rate of \citet{fakhouri_merger_2010}.

The magnitude of the time-redshift relation is 
\begin{equation}
\left|\frac{dt}{dz}\right| = \frac{1}{(1+z)H(z)}
\label{eq:dtdz}
\end{equation}
with $H(z)= H_0 \sqrt{\Omega_{\rm m}(1+z)^3+\Omega_\Lambda}$. Equation~\eqref{eq:dtdz} is evaluated with the Millennium cosmology on which the \citet{fakhouri_merger_2010} rate was calibrated, $\Omega_{\rm m}=0.25$, $\Omega_\Lambda=0.75$ and $h=0.73$, so that the merger rate and the time-redshift conversion refer to the same simulation. 

To draw the epoch of the next major merger backward in time from a starting redshift $z_{\rm start}$, we treat the merger sequence as a nonhomogeneous Poisson process whose intensity varies with halo mass and redshift. The cumulative hazard between $z_{\rm start}$ and $z$ is
\begin{equation}
\mathcal{H}(z\mid z_{\rm start})
=
\int_{z_{\rm start}}^z \frac{dN_{\rm major}}{dz'} dz'
\label{eq:hazard}
\end{equation}

The corresponding lookback time from $z_{\rm start}$ is
\begin{equation}
\Delta t(z\mid z_{\rm start}) = \int_{z_{\rm start}}^z \left|\frac{dt}{dz'}\right| dz'
\end{equation}
Both integrals are evaluated on a common grid of 5000 points in $z$, running from $z_{\rm start}$ to the redshift reached after the adopted lookback budget, capped at $z=10$.

For a Poisson process, the probability that no major merger occurs before the cumulative hazard reaches $\mathcal{H}$ is
\begin{equation}
P_0(\mathcal{H})=\exp(-\mathcal{H})
\end{equation}
We therefore draw a uniform random number $u_t\in(0,1]$ and define
\begin{equation}
E=-\ln u_t
\end{equation}
The next merger redshift is obtained by solving
\begin{equation}
\mathcal{H}(z_{\rm event}\mid z_{\rm start})=E
\end{equation}
If $E$ exceeds the maximum cumulative hazard reached within the adopted time interval, no major merger occurs within that interval and the realization is assigned no finite event time. Otherwise, inversion of the cumulative hazard gives $z_{\rm event}$, and
\begin{equation}
\Delta t_{\rm event}
=
\Delta t(z_{\rm event}\mid z_{\rm start})
\end{equation}
gives the corresponding lookback time to the merger.

\subsubsection{Most recent-shock selection}
\label{sec:recent_shock_selection}
For each descendant halo $M_0$ we draw the most recent major merger beginning at this halo's own $z_{0}$, within a maximum lookback budget $t_{\rm limit}=10\ {\rm Gyr}$. The resulting interval and redshift are
\begin{equation}
\Delta t_1 = t(z_{{\rm obs},0})-t(z_1), \qquad z_1 = z_{\rm first}
\end{equation}
The time limit for which we consider it the first merger is
$\Delta t_1 < \Delta t_{\rm shock,max}, \Delta t_{\rm shock,max} = 1\ {\rm Gyr}$. A ring can only be observed within its lifetime after the shock. The simulated ring lifetimes from the simulation are shorter than this window. Halos whose most recent major merger is older than $\Delta t_{\rm shock,max}$ therefore cannot host an observable ring, and we do not follow their bubbles. We draw descendant halos until $N_{\rm recent}=5\times10^4$ of them have $\Delta t_1<\Delta t_{\rm shock,max}$. This required $N_{\rm valid}=500{,}876$ draws with a first merger within the full window $t_{\rm limit}$, giving a recent-merger fraction
\begin{equation}
f_{\rm recent} = \frac{N_{\rm recent}}{N_{\rm valid}} = 0.0998
\end{equation}
This fraction is set by the merger rate (Equation~\eqref{eq:mergerrate}).

We continue drawing descendant halos until $N_{M_0}$ have passed this selection. Each accepted halo is also assigned a single shock Mach number at this point (Section~\ref{sec:shock_mach_number}), shared by all of its progenitor realizations.

\subsubsection{Progenitor mass decomposition and branch selection}
\label{sec:progenitor_mass_decomposition}
For every accepted first merger we draw a major-merger mass ratio $\xi$ and decompose the descendant mass into two progenitors summing to $M_0$,
\begin{equation}
M_{\rm large} = \frac{M_0}{1+\xi}, \qquad M_{\rm small} = \frac{\xi M_0}{1+\xi}
\end{equation}
One branch is followed backward to the preceding merger, selected with probability proportional to progenitor mass,
\begin{equation}
P_{\rm large} = \frac{1}{1+\xi}, \qquad P_{\rm small} = \frac{\xi}{1+\xi}
\end{equation}
The selected progenitor, $M_{\rm prog}$, is used to draw the preceding major merger of that same branch, beginning at $z_{\rm first}$ (not $z_{0}$) and searching within the remaining time budget $t_{\rm limit}-\Delta t_1$. This gives an interval $\Delta t_2$ and redshift $z_{\rm second}$; the total lookback time to the earlier merger is
\begin{equation}
\Delta t_{\rm total} = \Delta t_1 + \Delta t_2
\end{equation}
The line of history is rejected if $\Delta t_2$ is not finite, or if $\Delta t_{\rm total}\geq t_{\rm limit}$. This entire procedure is repeated $N_\xi=10$ times independently for each accepted descendant halo, giving $N_{M_0}\times N_\xi=$ total merger-history realizations, which we refer to as cases.

\begin{figure*}[t]
    \centering
    \includegraphics[width=0.5\textwidth]{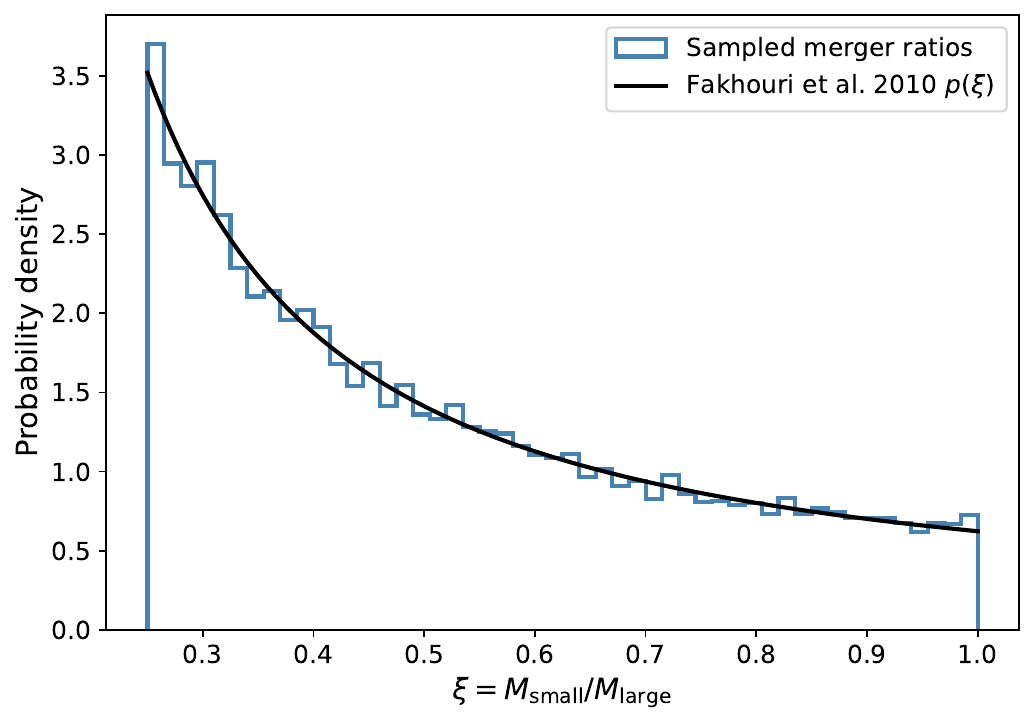}
    \caption{Distribution of sampled major-merger mass ratios $\xi$ (blue histogram) compared with the normalized probability density of Equation~\eqref{eq:pxi} from \citet{fakhouri_merger_2010} (black curve).}
    \label{fig:sampled_merger_ratios_vs_Fakhouri_PDF}
\end{figure*}

\subsubsection{Progenitor galaxy mass selection}
\label{sec:progenitor_galaxy_mass_selection}\label{sec:progmass}
The radio AGN host selected in Section~\ref{sec:stellar_mass_radioagn} with $M_\star$ is carried through the merger history unchanged. The progenitor mass $M_{\rm prog}$ therefore sets the galaxy's environment and its central/satellite classification. Following the notation of Section~\ref{sec:stellar_to_halo}, the conditional stellar mass function of \citet{yang_galaxy_2009} at $M_{\rm prog}$ splits into central and satellite terms, $\Phi_{\rm tot}(\log M_\star\mid M_{\rm prog})=\Phi_{\rm cen}+\Phi_{\rm sat}$, which we use to assign the central/satellite type at this galaxy's own $M_\star$:
\begin{equation}
P_{\rm cen} = \frac{\Phi_{\rm cen}(M_\star\mid M_{\rm prog})}{\Phi_{\rm cen}(M_\star\mid M_{\rm prog})+\Phi_{\rm sat}(M_\star\mid M_{\rm prog})}
\label{eq:pcen}
\end{equation}
A uniform random number $u_{\rm cen}$ classifies the galaxy as central if $u_{\rm cen}<P_{\rm cen}$, satellite otherwise. Centrals are placed at the halo center, $r_{\rm gal}=0$; satellites are assigned a nonzero halo-centric radius using the radial profile in the next section.

\subsubsection{Progenitor galaxy environment}
\label{sec:progenitor_galaxy_environment}\label{sec:progenv}\label{sec:progenitor_environment}
We first convert the progenitor halo mass into a halo radius,
\begin{equation}
M_{\rm prog} = \frac{4\pi}{3}R_{200{\rm m}}^3\left[200 \rho_{\rm m}(z_{\rm first})\right]
\label{eq:r200m_prog}
\end{equation}
where $\rho_{\rm m}(z)=\Omega_{\rm m}\rho_{\rm crit,0}(1+z)^3$ is the mean matter density at the epoch of the shock and $\rho_{\rm crit,0}=3H_0^2/8\pi G$. We evaluate the halo at $z_{\rm first}$, the redshift of the shock that produces the ring, since $M_{\rm prog}$ is by construction the progenitor mass at that merger.

We model the spatial number-density distribution of satellite galaxies with a Navarro-Frenk-White profile. Using stacked Sloan Digital Sky Survey systems, \citet{budzynski_radial_2012} found that the projected number-density profiles of satellite galaxies are broadly consistent with NFW-like distributions, with a fitted galaxy concentration of approximately $c_{\rm gal}=2.6$. The three-dimensional NFW number-density profile is
\begin{equation}
n_{\rm sat}(r) \propto \frac{1}{(r/r_{\rm s})(1+r/r_{\rm s})^2}, \qquad r_{\rm s} = \frac{R_{200{\rm m}}}{c_{\rm gal}}
\label{eq:nfw}
\end{equation}

The number of satellites in a shell of radius $r$ is $n_{\rm sat}(r) 4\pi r^2dr$, so with $x=r/r_{\rm s}$ the radial probability density is $p(x)\propto x/(1+x)^2$ and the enclosed number of satellites is proportional to $g(x)=\ln(1+x)-x/(1+x)$. We allow satellites to occupy the halo's outer infall region as well as its virialized interior, $x_{\rm max}=2c_{\rm gal}$, i.e. $0<r_{\rm gal}<2R_{200{\rm m}}$, so the normalized cumulative distribution is
\begin{equation}
F_{\rm sat}(x) = \frac{\ln(1+x)-x/(1+x)}{\ln(1+x_{\rm max})-x_{\rm max}/(1+x_{\rm max})}, \qquad 0\leq x\leq x_{\rm max}
\label{eq:fsat}
\end{equation}
For each satellite we draw a uniform random number $u_r\in[0,1]$, invert $F_{\rm sat}(x_{\rm draw})=u_r$, and set $r_{\rm gal}=x_{\rm draw} r_{\rm s}$. Extending the profile to $2R_{200{\rm m}}$ takes it beyond the radius over which it was measured, and we return to this in Section~\ref{sec:caveats}.

We assume the halo gas is approximately in virial equilibrium with the dark matter potential; the characteristic velocity dispersion is
\begin{equation}
\sigma_v^2 = \frac{GM_{\rm prog}}{2R_{200{\rm m}}}
\label{eq:sigmav}
\end{equation}
and the gas temperature follows from equating the thermal energy per particle to this characteristic kinetic energy,
\begin{equation}
k_{\rm B}T = \mu m_p \sigma_v^2 = \frac{\mu m_p G M_{\rm prog}}{2R_{200{\rm m}}}
\label{eq:temperature}
\end{equation}
with $\mu=0.59$ the mean molecular weight per free particle. We represent the halo density with a spherical beta model,
\begin{equation}
n_e(r) = n_{e,0}\left[1+\left(\frac{r}{r_c}\right)^2\right]^{-3\beta_{\rm gas}/2}
\label{eq:betamodel}
\end{equation}
with core radius $r_c=f_{\rm core}R_{200{\rm m}}$, $f_{\rm core}=0.1$ \citep{osullivan_heating_2011}, and $\beta_{\rm gas}=2/3$ \citep{cavaliere_x-rays_1976}. The gas mass implied by this profile,
\begin{equation}
M_{\rm gas} = 4\pi\mu_e m_p\int_0^{R_{200{\rm m}}} n_e(r)r^2 dr = \mu_e m_p n_{e,0}I_\beta,
\qquad
I_\beta = 4\pi\int_0^{R_{200{\rm m}}} r^2\left[1+\left(\frac{r}{r_c}\right)^2\right]^{-3\beta_{\rm gas}/2}dr,
\label{eq:Ibeta}
\end{equation}
with $\mu_e=1.17$ the mean molecular weight per free electron, fixes the central density. Substituting $r=x r_c$ makes the shape integral dimensionless,
\begin{equation}
I_\beta = 4\pi r_c^3 J,
\qquad
J = \int_0^{1/f_{\rm core}} x^2\left(1+x^2\right)^{-3\beta_{\rm gas}/2}dx
\label{eq:Jshape}
\end{equation}
so that $J$ depends only on $f_{\rm core}$ and $\beta_{\rm gas}$ and is evaluated once for the whole calculation. For $\beta_{\rm gas}=2/3$ it reduces to $J=f_{\rm core}^{-1}-\arctan(f_{\rm core}^{-1})=8.53$.

The hot gas mass is $M_{\rm gas}=f_{\rm gas}(M_{\rm prog})M_{\rm prog}$, where we adopt the mass-dependent hot gas fraction measured by \citet{popesso_hot_2026} from eROSITA/eFEDS stacking of optically selected galaxy groups,
\begin{equation}
f_{\rm gas}(M_{\rm prog}) = \min\left[2.09\times10^{-6}\left(\frac{M_{\rm prog}}{M_\odot}\right)^{0.33},\ \frac{\Omega_{\rm b}}{\Omega_{\rm m}}\right]
\label{eq:fgas}
\end{equation}
capped at the cosmic baryon fraction, which the powerlaw would exceed above $\log_{10}(M_{\rm prog}/M_\odot)\approx14.9$, outside the range over which it was fitted. At the group masses that dominate this calculation $f_{\rm gas}\approx0.02$-$0.04$. The central density is then
\begin{equation}
n_{e,0} = \frac{f_{\rm gas}(M_{\rm prog}) M_{\rm prog}}{\mu_e m_p I_\beta}
= \frac{f_{\rm gas}(M_{\rm prog}) \Delta_{\rm m} \rho_{\rm m}(z_{\rm first})}{3 \mu_e m_p f_{\rm core}^3 J}
\label{eq:ne0}
\end{equation}
where the second form follows from substituting Equation~\eqref{eq:r200m_prog}, and shows that the halo radius cancels. 

The electron density at any halo-centric radius $r$ follows from the beta model,
\begin{equation}
    n_e(r)=n_{e,0}\left[1+(r_{\rm gal}/r_c)^2\right]^{-3\beta_{\rm gas}/2}
\label{eq:negal}
\end{equation}
reducing to $n_e(r_{\rm gal})=n_{e,0}$ for centrals. Using $P_{\rm env}=n_{\rm tot}k_{\rm B}T$ with $n_{\rm tot}=(\mu_e/\mu)n_e$, the environmental pressure at the galaxy's position is
\begin{equation}
P_{\rm env}(r) = \frac{\mu_e}{\mu}n_e(r_{\rm gal}) k_{\rm B}T
\label{eq:penv}
\end{equation}
At the host galaxy, $r=r_{\rm gal}$, which reduces to $n_{\rm e,0}$ for centrals. We use $P_{\rm env}(r_{\rm gal})$ to decide which AGN episodes feed the same bubble (Section~\ref{sec:episodes}), and $P_{\rm env}$ at the bubble's own position for its size and for the ring.

We verified that the resulting host halos follow the observed cluster mass-temperature relation \cite{lovisari_scaling_2022}, with temperatures tracking the expected virial scaling $k_{\rm B}T\propto M_{\rm h}^{2/3}$.

\subsubsection{AGN episodes and bubble energies}
\label{sec:AGN_episodes_bubbles}\label{sec:episodes}\label{sec:bubbles}\label{sec:bubble_energies}
The AGN sequence for this realization is generated within the lookback-time interval bounded by the two sampled mergers, $\Delta t_1\leq t_{\rm lookback}<\Delta t_{\rm total}$. AGN duty cycles are not well measured or understood. A natural timescale to consider for individual AGN episodes is the black hole's Salpeter time, the time it would take a black hole to double its mass if it were accreting at Eddington, but many other choices are possible. Below, we adopt this as our fiducial model for the characteristic lifetime of individual AGN episodes.

Each episode's lifetime is then drawn independently from a log-normal distribution with its mean equal to one Salpeter time, $\langle t_{\rm jet}\rangle=45\ {\rm Myr}$, and variation $\sigma(t_{\rm jet})/\langle t_{\rm jet}\rangle=1$. Writing the log-normal in terms of these parameters,
\begin{equation}
\sigma_{\ln t}^2 = \ln\!\left(1+\left[\frac{\sigma(t_{\rm jet})}{\langle t_{\rm jet}\rangle}\right]^2\right), \qquad
\mu_{\ln t} = \ln\langle t_{\rm jet}\rangle - \tfrac{1}{2}\sigma_{\ln t}^2
\end{equation}

The most recent episode is allowed to overlap the most recent merger time $\Delta t_1$, but it is truncated at the shock: only the portion inflated before the shock passes contributes to the ring. For a drawn lifetime $t_{{\rm jet},1}$ and uniform overlap variable $u_1\sim\mathcal{U}(0,1)$,
\begin{equation}
t_{{\rm start},1} = \Delta t_1 + u_1 t_{{\rm jet},1}, \qquad
t_{{\rm end},1} = \Delta t_1,
\label{eq:episode1}
\end{equation}
so that only the pre-shock duration $u_1 t_{{\rm jet},1}$ is retained. This guarantees that no bubble is still being inflated after the shock has passed, i.e. the coast time of Equation~\eqref{eq:coast} is non-negative.

Subsequent episodes are populated contiguously going backward in time. Let episode $j-1$ be the previously accepted, more recent episode. A new lifetime $t_{{\rm jet},j}$ is drawn from the same log-normal, and the episode is placed so that it ends where the previous one began,
\begin{equation}
t_{{\rm end},j} = t_{{\rm start},j-1}, \qquad
t_{{\rm start},j} = t_{{\rm end},j} + t_{{\rm jet},j}
\label{eq:nextepisode}
\end{equation}
The AGN is therefore taken to be active throughout the interval between the two mergers, with the duty cycle divided into consecutive episodes of independently drawn duration and jet power. The sequence terminates when a proposed start time reaches the earlier-merger boundary, $t_{{\rm start},j}\geq\Delta t_{\rm total}$, with a maximum of $10^4$ attempted episodes per realization. Each accepted episode's lookback time is converted to redshifts $z_{{\rm start},j}$, $z_{{\rm end},j}$, and is assigned an independent kinetic Eddington ratio using the same radio-AGN incidence relation introduced in Section~\ref{sec:stellar_mass_radioagn} from eq.~(\ref{eq:oshea}), over the same range $-6\leq\log_{10}\lambda_{\rm jet}\leq-1$.

The central black-hole mass is estimated from the realization's fixed progenitor stellar mass, $M_{\rm BH}=0.002M_\star$ \citep[following the assumptions in][]{igo_global_2025}, giving an Eddington luminosity
\begin{equation}
L_{\rm Edd} = 1.26\times10^{38}\left(\frac{M_{\rm BH}}{M_\odot}\right)\ {\rm erg\ s^{-1}}
\end{equation}
The same $M_{\rm BH}$ and $L_{\rm Edd}$ apply to every episode in one realization, neglecting any black hole growth (a reasonable assumption for radio AGN, which generally accrete at low rates.) 
The kinetic power assigned to episode $j$,
\begin{equation}
Q_{{\rm jet},j} = \lambda_{{\rm jet},j}L_{\rm Edd}
\end{equation}
varies between episodes only through the independently-drawn $\lambda_{{\rm jet},j}$. The energy supplied during episode $j$, assuming constant jet power over its sampled lifetime, is
\begin{equation}
E_{{\rm jet},j} = Q_{{\rm jet},j} t_{{\rm jet},j}
\end{equation}
and in the current model all of this energy is assigned to inflating the radio bubble, $E_{{\rm bub},j}=E_{{\rm jet},j}$.

For a bubble dominated by relativistic particles (adiabatic index $\gamma_{\rm bub}=4/3$), the injected bubble energy is $E_{\rm bub}=4P_{\rm env}V_{\rm bub}$; approximating the bubble as a sphere of volume $V_{\rm bub}=\frac{4\pi}{3}R_{\rm bub}^3$,
\begin{equation}
E_{\rm bub} = \frac{16\pi}{3}P_{\rm env}R_{\rm bub}^3
\qquad\Longrightarrow\qquad
R_{\rm bub} = \left(\frac{3E_{\rm bub}}{16\pi P_{\rm env}}\right)^{1/3}
\label{eq:rbub}
\end{equation}
where $P_{\rm env}$ is the ambient pressure at the bubble's position, \textcolor{blue}{which we will determine below.} Equation~\eqref{eq:rbub} is the pressure-balance radius when the bubble reaches equilibrium with its surroundings during the late stage of its growing. During the early, strongly supersonic phase, the injected energy is carried by the driven shock, so $E_{\rm bub}=4P_{\rm env}V_{\rm bub}$ underestimates the energy at early times.

A single episode of typical duration and power inflates a bubble of only a few kpc, far smaller than the several hundred kpc structures observed as ORCs. However, successive outbursts along the same jet axis can re-enter and re-energize pre-existing lobes  \citep{kaiser_radio_2000}. We therefore allow consecutive episodes to deposit their energy into a common cavity. Grouping is performed forward in time, from the oldest episode toward the shock. After $n$ consecutive episodes have fed the same bubble,
\begin{equation}
E_{\rm acc} = \sum_{j=1}^{n} Q_{{\rm jet},j} t_{{\rm jet},j},
\qquad
t_{\rm acc} = \sum_{j=1}^{n} t_{{\rm jet},j}
\label{eq:accumulate}
\end{equation}

When the bubble is still expanding supersonically, the radius is described by the energy-driven wind solution \citep{weaver_interstellar_1977},
\begin{equation}
R_{\rm wind} = C_{\rm w}\left(\frac{L t_{\rm acc}^3}{\rho}\right)^{1/5},
\qquad C_{\rm w}=\left(\frac{250}{308\pi}\right)^{1/5}\approx0.76,
\label{eq:rwind}
\end{equation}
with mean mechanical power $L=E_{\rm acc}/t_{\rm acc}$ and ambient mass density
$\rho=\mu_e m_p n_e$. We take the physical radius to be the smaller of the two,
\begin{equation}
R_{\rm bub} = \min\!\left[R_{\rm wind},\ R_{\rm pb}(E_{\rm acc})\right],
\label{eq:rbub_switch}
\end{equation}
which switches continuously from the wind solution to pressure balance as the expansion velocity drops through the ambient sound speed.

We assume each AGN activity inflates a single, spherical cavity (bubble). A bipolar jet will generally inflate two lobes and could thus be producing the double-ring morphology seen in some ORCs (with correspondingly smaller radii, given the partition of the jet power between two bubbles.) We will discuss this in more detail in Section~\ref{sec:caveats}.

The chain ends when the host galaxy leaves the cavity it is feeding. The AGN separates from its cavity through two independent motions: the host galaxy's drift through the halo gas, and the buoyant rise of the bubble out of its own cavity. The host drift speed is set by the virial velocity dispersion of eq.~\eqref{eq:sigmav},
\begin{equation}
v_{\rm drift} = f_{v{\rm AGN}} \sigma_v,
\qquad
\sigma_v = \sqrt{\frac{k_{\rm B}T}{\mu m_p}},
\end{equation}
with $f_{v{\rm AGN}}=1$ for satellites and $f_{v{\rm AGN,cen}}=0.3$ for centrals; the three velocity components are drawn from a Maxwellian with per-axis dispersion $f_{v{\rm AGN}}\sigma_v$. A buoyant bubble rises out of its cavity at a fraction of the ambient sound speed \citep{churazov_evolution_2001},
\begin{equation}
v_{\rm buoy} = f_{\rm buoy} c_s,
\qquad
c_s = \sqrt{\frac{\gamma_{\rm gas} k_{\rm B}T}{\mu m_p}},
\end{equation}
with $\gamma_{\rm gas}=5/3$ and $f_{\rm buoy}=1.0$. Because drift and buoyant rise act in independent directions, the effective separation speed adds them in quadrature,
\begin{equation}
v_{\rm AGN} = \sqrt{v_{\rm drift}^2 + v_{\rm buoy}^2},
\label{eq:vagn}
\end{equation}
so that even a central galaxy at rest ($v_{\rm drift}\to0$) still leaves its cavity through buoyancy alone. The current episode is terminated when the accumulated displacement exceeds its radius,
\begin{equation}
v_{\rm AGN} t_{\rm acc} > R_{\rm bub}(E_{\rm acc})
\label{eq:residence}
\end{equation}
after which the next episode begins a new bubble. \textcolor{blue}{In this grouping step $R_{\rm bub}$ is evaluated with the pressure at the host galaxy, $P_{\rm env}(r_{\rm gal})$. However, a jet-inflated bubble is not concentric with the host galaxy itself. To model jet-inflated bubbles, we place its center at a distance equal to its own radius from the host galaxy, along a random direction $\hat{n}$ that is at an angle $\psi$ with the host's radius vector, with cos$\psi$ drawn uniformly in [-1,1]. The bubble center then is located at}
\begin{equation}
r_{\rm inf} = \left(r_{\rm gal}^2 + R_{\rm inf}^2 + 2\,r_{\rm gal}R_{\rm inf}\cos\psi\right)^{1/2}.
\label{eq:rinf}
\end{equation}
\textcolor{blue}{For central galaxies, the bubble center is simply $r_{\rm inf}=R_{\rm inf}$. The bubble radius then depends on the environment at this position}
\begin{equation}
R_{\rm inf} = R_{\rm bub}\!\left[E_{\rm acc},\,t_{\rm acc},\,P_{\rm env}(r_{\rm inf}),\,n_{\rm e}(r_{\rm inf})\right],
\label{eq:Rinf}
\end{equation}
\textcolor{blue}{with $R_{\rm bub}$ from Equation~\eqref{eq:rbub_switch}. We iterate Equation~\eqref{eq:Rinf} starting from the radius created at the host galaxy until $R_{\rm inf}$ the fractional change is less than $10^{-4}$.}

After the last episode inflating the bubble ends, the bubble waits until the shock arrives. We define the coast time between the jet switching off and the arrival of the shock
\begin{equation}
t_{\rm coast} = t_{{\rm end}, {\rm youngest}} - \Delta t_1
\label{eq:coast}
\end{equation}
where $t_{{\rm end}, {\rm youngest}}$ is the lookback time at which the most recent episode feeding the bubble ended. \textcolor{red}{During this time, the bubble rises buoyantly at $v_{\rm buoy}$. Buoyancy points away from the center of the halo potential, so the bubble moves radially outward from its inflation position, and when the shock arrives its center lies at}
\begin{equation}
\textcolor{red}{r_{\rm shock} = r_{\rm inf} + v_{\rm buoy}t_{\rm coast}}
\label{eq:rshock}
\end{equation}
\textcolor{red}{from the halo center. For central galaxies the radial direction is $\hat{n}$ itself, so the bubble keeps moving along the direction in which it was inflated.} \textcolor{blue}{It expands adiabatically into the lower pressure regions, so the radius of a bubble right before the shock hits becomes}
\begin{equation}
R_{\rm shock} = R_{\rm inf}\left[\frac{P_{\rm env}(r_{\rm inf})}{P_{\rm env}(r_{\rm shock})}\right]^{1/(3\gamma_{\rm bub})}
= R_{\rm inf}\left[\frac{P_{\rm env}(r_{\rm inf})}{P_{\rm env}(r_{\rm shock})}\right]^{1/4}.
\label{eq:Rshock}
\end{equation}
\textcolor{blue}{The ring produced by the shock is then computed from the bubble radius $R_{\rm shock}$ and the environment at $r_{\rm shock}$, described in later section (Section~\ref{sec:flux_grid}).}

A fossil bubble does not persist indefinitely. Buoyant rise, Rayleigh-Taylor, and Kelvin-Helmholtz instabilities can disrupt it and mix its contents into the ambient gas, so a bubble that formed long before the shock arrived is no longer a coherent structure that can be re-energized into a ring. \textcolor{blue}{We require that the bubble does not grow by more than $f_{\rm survive}$ times its inflated radius before the shock arrives,}
\begin{equation}
t_{\rm coast} \leq t_{\rm survive},\qquad
t_{\rm survive} = \min\left(f_{\rm survive}\frac{R_{\rm inf}}{v_{\rm buoy}},\ t_{\rm survive,max}\right),
\label{eq:tsurvive}
\end{equation}
with $t_{\rm survive,max}=10\ {\rm Gyr}$ is a wide safety ceiling, much longer than any reasonable survival time. We take $f_{\rm survive}=3$ as our fiducial value. \textcolor{blue}{The surviving bubbles then sit between $R_{\rm inf}$ and $4R_{\rm inf}$ from their host when the shock arrives.}
Bubbles with $t_{\rm coast}>t_{\rm survive}$ are considered dissipated and removed from the sample.

\begin{figure*}[t]
    \centering
    \includegraphics[width=\textwidth]{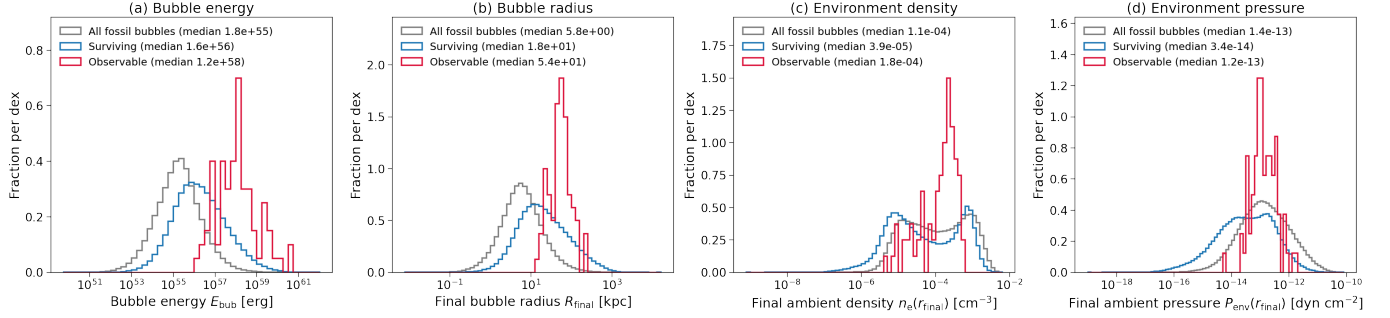}
    \caption{Normalized distributions (probability density per dex) of (a) bubble energy $E_{\rm bub}$, ( (b) final bubble radius $R_{\rm final}$, (c) the ambient electron density $n_{\rm e}$ and (d) the ambient pressure $P_{\rm env}$ at the bubble's final position, for all fossil bubbles (grey), the surviving bubbles (blue), and the observable subset that passes all survey cuts (red). The final state is the bubble just before the shock reaches it or, for bubbles that dissipate first, when it has risen $f_{\rm survive}R_{\rm inf}$ beyond its inflation position (Equation~\eqref{eq:tsurvive}); the radius includes the adiabatic expansion during the rise (Equation~\eqref{eq:Rshock}). The two peaks in panels (c) and (d) correspond to bubbles of central galaxies, which remain near the dense halo core, and bubbles of satellites, which sit at larger radii in much lower-density gas. Median values are given in the legends.}
    \label{fig:bubble_energy_radius_density}
\end{figure*}

Figure~\ref{fig:bubble_energy_radius_density} shows the bubble energy $E_{\rm bub}$, \textcolor{blue}{its radius $R_{\rm final}$,  the ambient density, and the ambient pressure at its final position} for the surviving bubbles and for the subset that pass all survey cuts (observable) (Section~\ref{sec:survey}). The surviving bubbles have energies near $\sim\textcolor{red}{1.59\times10^{56}} {\rm erg}$ and radii of a few tens of kpc (median $R_{\rm bub}\approx\textcolor{red}{18.50} {\rm kpc}$). The observable bubbles peak roughly two orders of magnitude higher in energy ($\sim \textcolor{red}{1.01\times10^{58}} {\rm erg}$) and at $R_{\rm bub}\approx \textcolor{red}{56.10} {\rm kpc}$ (diameters of $\sim\textcolor{red}{112.20} {\rm kpc}$). Surviving bubbles are located in lower density regions with median density $\textcolor{red}{3.62\times 10^{-5}}\,{\rm cm^{-3}}$ and pressure $\textcolor{red}{3.22\times 10^{-14}}\,{\rm dyn\,cm^{-2}}$. Bubbles that make observable rings later however are in not so less dense region $\textcolor{red}{1.50\times 10^{-4}}\,{\rm cm^{-3}}$ and pressure $\textcolor{red}{1.02\times 10^{-13}}\,{\rm dyn\,cm^{-2}}$. A ring must be large enough to be resolved and bright enough to be detected. In the dense cores the high pressure keeps bubbles of a given energy too small to pass the angular-size cut, while in the most dilute gas the rings are too faint, since their surface brightness rises steeply with the ambient pressure. The detectable bubbles therefore lie between the two peaks of the density distribution.

\begin{figure*}[t]
    \centering
    \includegraphics[width=\textwidth]{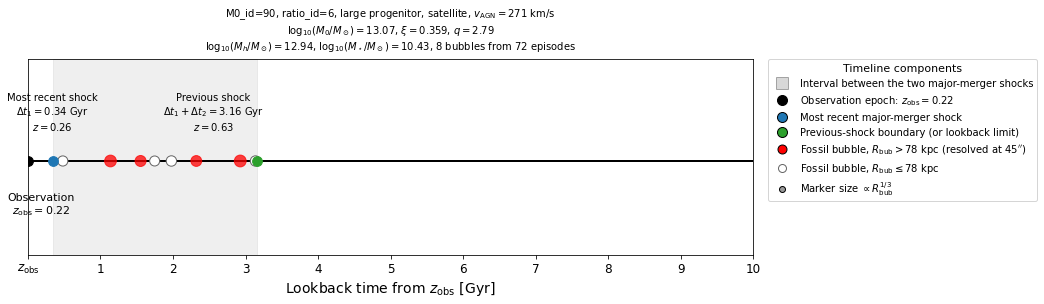}
    \caption{Progenitor galaxies and AGN episodes generated between the two sampled major-merger events. The starting time is the time of observation $z_{\rm obs}$ in black. The blue point is the most recent shock time by selecting $\delta t_1$ from the observation time and looking backward. The green point is the previous shock time by selecting $\delta t_2$ looking backward from the previous shock time. The blue and red points are AGN episode starting times by selecting $t_{\rm start, j}$. The red points are AGN episodes with $R_{\rm bub}>100$ kpc.}
    \label{fig:progenitor_galaxies_agn_episodes}
\end{figure*}

Figure~\ref{fig:progenitor_galaxies_agn_episodes} shows one example realization: the single $(\xi, M_{\rm prog})$ branch. Each point marks one AGN episode's bubble radius against its lookback time, illustrating how a chain of many short-lived episodes fills the interval between the two merger shocks $\Delta t_1$ and $\Delta t_{\rm total}$ for a single progenitor galaxy.

\subsubsection{Bubble weighting}
\label{sec:bubble_weighting}
The selected radio-AGN density from Section~\ref{sec:stellar_mass_radioagn} is
\begin{equation}
n_{\rm radioAGN}
= N_{\rm AGN}w_{\rm gal}
\end{equation}
where $w_{\rm gal}$ is the comoving number-density weight of one parent Monte Carlo galaxy.

The comoving density of selected radio AGN hosted by halos with a sufficiently recent major merger is
\begin{equation}
n_{\rm recentAGN}
= f_{\rm recent}  n_{\rm radioAGN}
\label{eq:nrecent}
\end{equation}

Let $w_{{\rm case},k}$ be the number-density weight of case $k$, obtained by dividing $n_{\rm recentAGN}$ evenly across all simulated cases. Each merger shock sweeps through the whole halo and encounters every fossil bubble still present, and the observation of ORCs with more than one concentric ring \citep{gasperin_single-_2026} indicates that multiple bubbles around the same galaxy can be visible simultaneously. We therefore treat every surviving bubble as an independent ORC candidate, assigning each the full weight of its parent case,
\begin{equation}
w_{{\rm bub},j} = w_{{\rm case},k}
\qquad\Longrightarrow\qquad
\sum_{j\in k} w_{{\rm bub},j} = n_{k} w_{{\rm case},k}
\label{eq:wbub}
\end{equation}
where $n_k$ is the number of bubbles in case $k$ that survive the criterion of Equation~\eqref{eq:tsurvive}. 

\begin{table*}[t]
\centering
\footnotesize
\setlength{\tabcolsep}{4pt}
\caption{Parameters of the population synthesis calculation. Values are those adopted in the fiducial ASKAP-EMU Pilot run.}
\label{tab:parameters}
\begin{tabular}{@{}l l >{\raggedright\arraybackslash}p{0.38\textwidth} >{\raggedright\arraybackslash}p{0.26\textwidth}@{}}
\hline\hline
Symbol & Value & Description & Reference \\
\hline
\multicolumn{4}{l}{\textit{Monte Carlo sample sizes}}\\
$N_{\rm gal}$              & $10^{6}$        & Sampled galaxy stellar masses & Sect.~\ref{sec:stellar_mass_radioagn}; \citet{bernardi_stellar_2018}\\
$N_{M_0}$                  & $5\times10^{4}$ & Accepted recent-shock descendant halos & Sect.~\ref{sec:recent_shock_selection}; \citet{yang_galaxy_2009}\\
$N_\xi$                    & 10              & Merger-history realizations per halo & Sect.~\ref{sec:progenitor_mass_decomposition}; \citet{fakhouri_merger_2010}\\
$N_{M_0}\times N_\xi$      & $5\times10^{5}$ & Total cases & \\
--                         & $10^{4}$        & Maximum attempted episodes per case & \\
\hline
\multicolumn{4}{l}{\textit{Comoving number densities and weights}}\\
$n_{\rm gal}$              & $1.30\times10^{-2}$ Mpc$^{-3}$ & Parent galaxy density & \citet{bernardi_stellar_2018}\\
$w_{\rm gal}$              & $1.30\times10^{-8}$ Mpc$^{-3}$ & Weight per sampled galaxy, Eq.~\eqref{eq:wgal} & \citet{bernardi_stellar_2018}\\
$n_{\rm radioAGN}$         & $1.83\times10^{-3}$ Mpc$^{-3}$ & Selected radio-AGN density & \citet{igo_lofar_2024,oshea_shooting_2025}\\
$f_{\rm recent}$           & 0.099           & Fraction with $\Delta t_1<\Delta t_{\rm shock,max}$ & \citet{fakhouri_merger_2010}\\
$n_{\rm recentAGN}$        & $1.80\times10^{-4}$ Mpc$^{-3}$ & Recent-shock radio-AGN density & \\
$w_{{\rm case},k}$         & $n_{\rm recentAGN}/N_{\rm case}$ & Weight per case & \\
$w_{{\rm bub},j}$          & $w_{{\rm case},k}$ & Weight per surviving bubble, Eq.~\eqref{eq:wbub} & \\
\hline
\multicolumn{4}{l}{\textit{AGN and jet}}\\
$a_\eta, b_\eta, c_\eta$ & $1.55, -0.845, -21.7$ & Radio-AGN incidence plane & \citet{oshea_shooting_2025}\\
$\log_{10}\lambda_{\rm jet}$ & $[-6,-1]$     & Kinetic Eddington ratio range & \citet{igo_lofar_2024}\\
$f_{\rm BH}$               & 0.002           & $M_{\rm BH}=f_{\rm BH}M_\star$ & \citet{marconi_relation_2003,igo_global_2025}\\
$\langle t_{\rm jet}\rangle$ & 45 Myr        & Mean episode lifetime (Salpeter time) & \citet{merloni_synthesis_2008}\\
$\sigma(t_{\rm jet})/\langle t_{\rm jet}\rangle$ & 1.0 & Coefficient of variation & \\
\hline
\multicolumn{4}{l}{\textit{Halo gas and galaxy position}}\\
$\Delta_{\rm m}$           & 200             & Overdensity, $M_{200{\rm m}}$ definition & \citet{tinker_toward_2008}\\
$\beta_{\rm gas}$          & 2/3             & Beta-model slope & \\
$f_{\rm core}$             & 0.1             & Core radius, $r_c=f_{\rm core}R_{200{\rm m}}$ & \\
$c_{\rm gal}$              & 2.6             & Satellite NFW concentration & \citet{budzynski_radial_2012}\\
$x_{\rm max}$              & $2c_{\rm gal}$  & Satellites out to $2R_{200{\rm m}}$ & \citet{budzynski_radial_2012}\\
$\mu,\ \mu_e$              & 0.59,\ 1.17     & Mean mass per particle, per electron & \\
$f_{\rm gas}(M)$           & $2.09\times10^{-6}M^{0.33}$ & Hot gas fraction, capped at $\Omega_{\rm b}/\Omega_{\rm m}$ & \citet{popesso_hot_2026}\\
\hline
\multicolumn{4}{l}{\textit{Merger and shock}}\\
$t_{\rm limit}$            & 10 Gyr          & Maximum lookback budget & \\
$\Delta t_{\rm shock,max}$ & 1 Gyr           & Recent-shock selection & \citet{ha_properties_2018,lee_morphology_2024}\\
$\xi_{\min}, \xi_{\max}$  & 0.25,\ 1.0      & Major-merger mass ratio range & \citet{lee_morphology_2024}\\
$\mathcal{M}_{\rm peak}$   & 3.0             & Mode of the Mach distribution & \citet{lee_morphology_2024}\\
$\sigma_{\ln\mathcal{M}}/\mu_{\ln\mathcal{M}}$ & 0.33 & Mach scatter ratio & \citet{lee_morphology_2024}\\
\hline
\multicolumn{4}{l}{\textit{Bubble growth and survival}}\\
$C_{\rm w}$                & 0.76            & Wind-bubble coefficient $(250/308\pi)^{1/5}$, Eq.~\eqref{eq:rwind} & \citet{weaver_interstellar_1977}\\
$f_{v{\rm AGN}}$           & 1.0             & Host drift, $v_{\rm drift}=f_{v{\rm AGN}}\sigma_v$ (satellites) & \\
$f_{v{\rm AGN,cen}}$       & 0.3             & Central-galaxy drift fraction & \\
$f_{\rm buoy}$             & 1.0             & Buoyant rise, $v_{\rm buoy}=f_{\rm buoy}c_s$, Eq.~\eqref{eq:vagn} & \citet{churazov_evolution_2001}\\
$t_{\rm survive,max}$      & 10 Gyr          & Coherence-time ceiling, Eq.~\eqref{eq:tsurvive} & \\
\hline
\multicolumn{4}{l}{\textit{Survey selection}}\\
$A_{\rm survey}$           & 270 deg$^2$     & Survey area & \citet{norris_unexpected_2021,norris_evolutionary_2021}\\
$z_{\min}, z_{\max}$      & 0.01,\ 0.55     & Survey redshift range & \citet{norris_odd_2021}\\
$V_{\rm survey}$           & $2.24\times10^{8}$ Mpc$^3$ & Effective comoving volume & \citet{diemer_colossus_2018}\\
$\theta_{\rm beam}$        & 11.96 arcsec    & $\sqrt{13''\times11''}$, Eq.~\eqref{eq:beam} & \citet{norris_evolutionary_2021}\\
$\theta_{\rm diam}$        & 45--600 arcsec  & Angular-diameter cut (ring) & \citet{norris_unexpected_2021}\\
$R_{\rm ring}/R_{\rm bub}$ & 1.0--1.2        & Ring-to-bubble radius, measured from grid & \citet{wang_odd_2026}\\
$R_{\rm ring}/w$           & $\sim$2         & Ring radius-to-width, measured from grid, Eq.~\eqref{eq:nbeam} & \citet{wang_odd_2026}\\
$\sigma_{\rm rms}$         & 0.025 mJy beam$^{-1}$ & Image noise & \citet{norris_evolutionary_2021}\\
$N_\sigma$                 & 5               & Detection significance & \\
$S_{\rm cut}$              & 0.125 mJy       & Flux and surface-brightness threshold & \\
$\nu_{\rm obs}$            & 944 MHz         & Observing frequency & \citet{norris_evolutionary_2021}\\
$\beta_{\rm plasma}$       & 10              & Bubble plasma beta & \citet{wang_odd_2026}\\
$\mathcal{M}$ grid         & 1.67-16           & Flux-grid Mach coverage & \citet{wang_odd_2026}\\
$P_{\rm orientation}$      & 0.5             & Viewing-angle probability & \citet{wang_odd_2026}\\
\hline
\multicolumn{4}{l}{\textit{Numerical grids}}\\
Stellar mass function      & 5000 pts, $9.05$--$12.25$ & Interpolation of Table B2 & \citet{bernardi_stellar_2018}\\
Halo mass (CSMF)           & 1000 pts, $12.16$--$14.58$ & Table 4 range ($h^{-1}M_\odot$) & \citet{yang_galaxy_2009}\\
Merger mass ratio          & 5000 pts        & $p(\xi)$, Eq.~\eqref{eq:pxi} & \citet{fakhouri_merger_2010}\\
Cumulative hazard          & 5000 pts        & Per merger draw, Eq.~\eqref{eq:hazard} & \citet{fakhouri_merger_2010}\\
Time--redshift table       & 20000 pts, $z\leq10.5$ & Eq.~\eqref{eq:dtdz} & \\
Survey volume CDF          & 20000 pts       & Eq.~\eqref{eq:zdraw} & \\
Beta-model shape integral  & 2000 pts        & $J$, Eq.~\eqref{eq:Jshape} & \\
Progenitor stellar mass    & 3000 pts        & $8.0$ to $\min(\log M_{\rm h}-0.05, 12.25)$ & \citet{bernardi_stellar_2018}\\
Satellite NFW CDF          & 5000 pts        & $F_{\rm sat}(x)$, Eq.~\eqref{eq:fsat} & \citet{budzynski_radial_2012}\\
\hline
\end{tabular}
\end{table*}

\subsection{Shock mach number}
\label{sec:shock_mach}\label{sec:mach}\label{sec:shock_mach_number}
The shock populations studied by \citet{vazza_shock_2009} span a broad range of Mach numbers and environments. \citet{lee_morphology_2024} studied the morphology and Mach-number distribution of merger-shock surfaces in simulations of merging galaxy clusters. The shock forms an extended curved surface with a distribution of local Mach numbers represented by a log-normal distribution, peaking somewhere in the range $\mathcal{M}_{\rm peak}\approx2$-$4.5$ depending on the specific merger, with a ratio of the log-space scatter to the log-space mean, $\sigma_{\ln\mathcal{M}}/\mu_{\ln\mathcal{M}}$, that stays close to $0.33$ across their sample of merging cluster pairs. We adopt this same shape, fixing the peak of our distribution at the middle of their reported range, $\mathcal{M}_{\rm peak}=3.0$, and using their reported scatter ratio directly, $\frac{\sigma_{\ln\mathcal{M}}}{\mu_{\ln\mathcal{M}}}=0.33$. From their Figure 4, the mode of a log-normal distribution with parameters $\mu_{\ln\mathcal{M}}$ and $\sigma_{\ln\mathcal{M}}$ is $\mathcal{M}_{\rm peak}=\exp(\mu_{\ln\mathcal{M}}-\sigma_{\ln\mathcal{M}}^2)$, which for $\mathcal{M}_{\rm peak}=3.0$ gives $\mu_{\ln\mathcal{M}}=1.276$ and $\sigma_{\ln\mathcal{M}}=0.421$.

\begin{figure*}[t]
    \centering
    \includegraphics[width=0.5\textwidth]{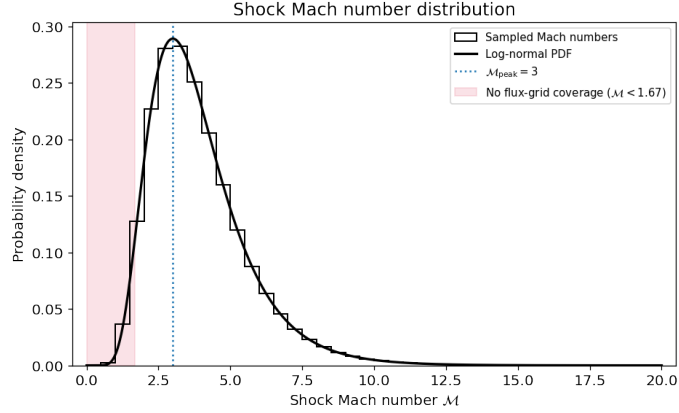}
    \caption{Distribution of shock Mach numbers, drawn independently for each fossil bubble from the local Mach-number distribution of the shock surface. The histogram shows the sampled distribution; the black curve is the log-normal probability density with $\mathcal{M}_{\rm peak}=3.0$ and $\sigma_{\ln\mathcal{M}}/\mu_{\ln\mathcal{M}}=0.33$ from \citet{lee_morphology_2024}. The shaded band marks $\mathcal{M}<1.67$ in the population but contributes zero flux due to not being able to form rings.}
    \label{fig:shock_mach}
\end{figure*}

Because the shock is an extended surface with varying strengths across it, different fossil bubbles can be hit by different parts of the same shock front and experience different local Mach numbers. We draw an independent Mach number from the log-normal for each bubble. All bubbles associated with a given shock sample the same underlying distribution but take distinct local values. Figure~\ref{fig:shock_mach} shows the resulting sampled distribution together with the underlying log-normal.

\textcolor{black}{The MHD flux grid used in Section~\ref{sec:survey_selection} spans
$1.67\leq\mathcal{M}\leq16$. Shocks with $\mathcal{M}<1.67$ are not expected to produce observable ORCs, and the grid is not extended below this value. Those rings in the population are assigned zero flux. We will discuss this more in the next section.}

\subsection{\textcolor{black}{Synthetic Flux and radius calculation}}
\label{sec:flux_grid}

The radio flux and the projected ring size from each bubble are computed from a pre-computed grid of MHD simulations of Shock-Bubble interactions \citep{wang_odd_2026}. A set of simulations was run at fixed shock Mach numbers ($\mathcal{M}=1.67, 2, 2.83, 4, 5.66, 8, 16$). The initial radius of the bubble is $R_0=10 {\rm kpc}$. The shock compresses it into an edge-brightened ring \ref{fig:orc1_observed_obs}. For each Mach number we record the snapshots labeled by ring age $t_{\rm ring}$ (time since shock passage), and produce synthetic radio images at 30 frequencies from $0.3 {\rm MHz}$ to $800 {\rm GHz}$, treating synchrotron and inverse-Compton cooling separately. We follow the scaling approach presented in \citet{wang_odd_2026}to allow us to calculate fluxes for each ring given its pressure, radius, redshift, and observing frequency.
To this end, each image is reduced to a luminosity at a fixed reference distance ($z=0.1065$) and a reference plasma
$\beta_0\equiv P_{\rm gas,0}/P_{\rm mag,0}$, giving
$L_\nu^{\rm synch}(\mathcal{M},t_{\rm ring},\nu)$,
$L_\nu^{\rm CMB}(\mathcal{M},t_{\rm ring},\nu)$, and $\beta_0(\mathcal{M},t_{\rm ring})$.
From each image, we also measure the ring peak-intensity radius $R_{\rm ring}$, the FWHM width $w$, and the aspect ratio $R_{\rm ring}/w$ at every $(\mathcal{M},t_{\rm ring},\nu)$.

\paragraph{Ring radius.}
The observed ring has a radius and width that evolve with ring age and differ from $R_{\rm bub}$. We keep the ring-radius-over-width ratio $R_{\rm ring}/w$ to the reference bubble and rescale them to each bubble's own radius
\begin{equation}
R_{\rm ring} = \frac{R_{\rm ring}}{R_0}\bigg|_{\mathcal{M},t_{\rm ring}} R_{\rm bub},
\qquad
w = \frac{R_{\rm ring}}{a}, \qquad
a \equiv \frac{R_{\rm ring}}{w}\bigg|_{\mathcal{M},t_{\rm ring}},
\label{eq:ring_scaling}
\end{equation}
where the ratio $R_{\rm ring}/R_0$ and aspect $a$ are taken at the bubble's Mach number and ring age. The ring radius $R_{\rm ring}$ sets the angular size (Section~\ref{sec:survey_selection}), and the aspect ratio $a$ sets the emitting annulus used for the surface brightness.

Each bubble's flux, spectral index, and ring geometry are obtained by logarithmically interpolating the grid in $(\mathcal{M},t_{\rm ring},\nu)$ and rescaling with Equations~\eqref{eq:lum_scaling} and \eqref{eq:ring_scaling}. Bubbles whose Mach number or ring age falls outside the grid coverage are assigned zero flux and no ring, and cannot pass the selection below. In particular, shocks with
$\mathcal{M}<2$ are too weak to re-energize the bubble into a bright, edge-brightened ring.

\paragraph{Flux.}
Each grid image provides a reference luminosity $L_{\nu,0}$ (integrating intensity over the image and calculating the corresponding luminosity at the reference distance) for each cooling channel at the reference bubble radius $R_0$, field $B_0$, and pressure $P_{\rm gas,0}$. A bubble with its own $R$, $B$, $P_{\rm gas}$, and redshift $z$ rescales
these by the dimensionless factor \citep{wang_odd_2026}
\begin{equation}
\label{eq:lum_scaling}
\mathcal{L} = \left(\frac{R}{R_0}\right)^3\left(\frac{B}{B_0}\right)^{3/2}
\left(\frac{P_{\rm gas}}{P_{\rm gas,0}}\right)\frac{1}{1+z},
\end{equation}
so that the rescaled luminosity in each channel is $L_\nu = \mathcal{L} L_{\nu,0}$. Radiative cooling steepens each spectrum above a cooling frequency. When rescaling, we shift the grid spectrum in frequency to the new cooling frequency, using the scale-free cooling method of \citet{wang_odd_2026}, to which we refer the reader for details.

The synchrotron and inverse-Compton channels are then combined
\begin{equation}
L_\nu^{\rm total} = \left[\left(L_\nu^{\rm CMB}\right)^{-1}
+ \left(L_\nu^{\rm synch}\right)^{-1}\right]^{-1},
\end{equation}
so that the more strongly cooled (fainter) channel limits the emergent luminosity. The observed flux density then follows from the bubble's luminosity distance,
\begin{equation}
S_\nu = \frac{L_\nu^{\rm total}}{4\pi D_L(z)^2}.
\end{equation}

\subsection{\textcolor{black}{Survey selection for ASKAP-EMU Pilot}}
\label{sec:survey_selection}\label{sec:survey}\label{sec:results}
The calculation above yields, for each candidate bubble, an observed flux density and a projected ring size, but the rings they produce are not necessarily detectable as Odd Radio Circles. Detectability depends on distance, angular size, flux density, surface brightness, and viewing angle. We apply a survey-selection model based on the ASKAP observations in which the first ORCs were discovered \citep{norris_unexpected_2021,norris_odd_2021}.

The adopted survey area is $A_{\rm survey}=270 {\rm deg^2}$, a fraction $f_{\rm sky}=A_{\rm survey}/41253 {\rm deg^2}$ of the sky. With redshift limits $0.01<z_{\rm obs}<0.55$, the effective comoving survey volume is $V_{\rm survey}=f_{\rm sky} [V_{\rm c}(z_{\max})-V_{\rm c}(z_{\min})] =2.24\times10^{8} {\rm Mpc^3}$. Each ring is placed at the observation redshift $z_{\rm obs}$.

The angular radius of a ring of physical radius $R_{\rm ring}$ (Equation~\ref{eq:ring_scaling}) at redshift $z$ is
\begin{equation}
\theta_{\rm ring} = \frac{R_{\rm ring}}{D_{\rm A}(z)},
\end{equation}
with $D_{\rm A}$ the angular-diameter distance and angular diameter $\theta_{\rm diam}=2\theta_{\rm ring}$. The ASKAP beam has
$\theta_{\rm beam,maj}=13 {\rm arcsec}$ and $\theta_{\rm beam,min}=11 {\rm arcsec}$. 
We use
\begin{equation}
\theta_{\rm beam}=\sqrt{\theta_{\rm beam,maj} \theta_{\rm beam,min}}\simeq11.96 {\rm arcsec}.
\label{eq:beam}
\end{equation}
The angular-diameter cut is $45 {\rm arcsec}<\theta_{\rm diam}<600 {\rm arcsec}$ ($3.8$-$50.2$ beam widths). The lower bound requires the ring to be resolved across several beams so that its edge-brightened morphology is recognizable. The upper bound lies well above the known ORC sizes and our population shows no ring can inflate to that size.

Taking the beam to be Gaussian, its solid angle is
\begin{equation}
\Omega_{\rm beam}=\frac{\pi}{4\ln 2} \theta_{\rm beam}^2\simeq1.133 \theta_{\rm beam}^2.
\label{eq:omegabeam}
\end{equation}
The ring emission fills an annulus of angular radius $\theta_{\rm ring}$ and aspect ratio $a$ (Equation~\ref{eq:ring_scaling}), of solid angle
$\Omega_{\rm source}=2\pi\theta_{\rm ring}^2/a$, so the number of beams over the source is
\begin{equation}
N_{\rm beam}=\max\!\left(\frac{\Omega_{\rm source}}{\Omega_{\rm beam}},\ 1\right),
\label{eq:nbeam}
\end{equation}
with the floor at unity for sources smaller than the beam.
For the EMU Pilot Survey we adopt an rms image noise of $\sigma_{\rm rms}=0.025\textcolor{red}{\,}{\rm mJy\textcolor{red}{\,}beam^{-1}}$, the lower end of the \textcolor{red}{25-30}\,$\mu{\rm Jy\textcolor{red}{\,}beam^{-1}}$ range reported by \citet{norris_evolutionary_2021}, and we require a $N_\sigma=5$ detection, giving a surface-brightness threshold
\begin{equation}
S_{\rm beam,cut}=N_\sigma\sigma_{\rm rms}=0.125\textcolor{red}{\,}{\rm mJy\textcolor{red}{\,}beam^{-1}}.
\label{eq:sbcut}
\end{equation}

The integrated flux density $S_\nu$ is spread over $N_{\rm beam}$ beams, so the surface brightness (flux per beam) is
\begin{equation}
S_{\nu,{\rm beam}} = \frac{S_\nu}{N_{\rm beam}}.
\label{eq:sbeam}
\end{equation}

\textcolor{red}{The noise summed over $N_{\rm beam}$ independent beams grows as $\sqrt{N_{\rm beam}}$, so the} corresponding threshold on $S_\nu$ is
\begin{equation}
S_{\rm total,cut}=N_\sigma\sigma_{\rm rms}\sqrt{N_{\rm beam}}.
\label{eq:stotcut}
\end{equation}
The two flux conditions are $S_\nu>S_{\rm total,cut}$ and $S_{\nu,{\rm beam}}>S_{\rm beam,cut}$.

Each bubble carries a comoving weight $w_{{\rm bub},j}$. Writing $\mathcal{S}_{\theta,j}$, \textcolor{red}{$\mathcal{S}_{S,j}$ and} $\mathcal{S}_{{\rm beam},j}$ for the angular-size, \textcolor{red}{total-flux} and surface-brightness selection functions, the expected numbers in the survey at successive stages are
\begin{align}
N_{\rm all}&=V_{\rm survey}\sum_j w_{{\rm bub},j},\\
N_{\theta}&=V_{\rm survey}\sum_j w_{{\rm bub},j} \mathcal{S}_{\theta,j},\\
\textcolor{red}{N_{S}}&\textcolor{red}{=V_{\rm survey}\sum_j w_{{\rm bub},j} \mathcal{S}_{\theta,j} \mathcal{S}_{S,j},}\\
N_{\rm detectable}&=V_{\rm survey}\sum_j w_{{\rm bub},j}
\mathcal{S}_{\theta,j} \textcolor{red}{\mathcal{S}_{S,j}} \mathcal{S}_{{\rm beam},j}.
\end{align}
A compressed bubble does not appear as a circular ring from every direction\textcolor{red}{. We} thus apply an orientation probability $P_{\rm orientation}=0.5$\textcolor{red}{, and} the expected number of ORC-like sources is
\begin{equation}
N_{\rm ORC}=P_{\rm orientation} N_{\rm detectable}.
\end{equation}
For the fiducial run this gives $N_{\rm ORC}\simeq \textcolor{red}{2.70}$.

\begin{figure*}[t]
    \centering
    \includegraphics[width=0.5\textwidth]{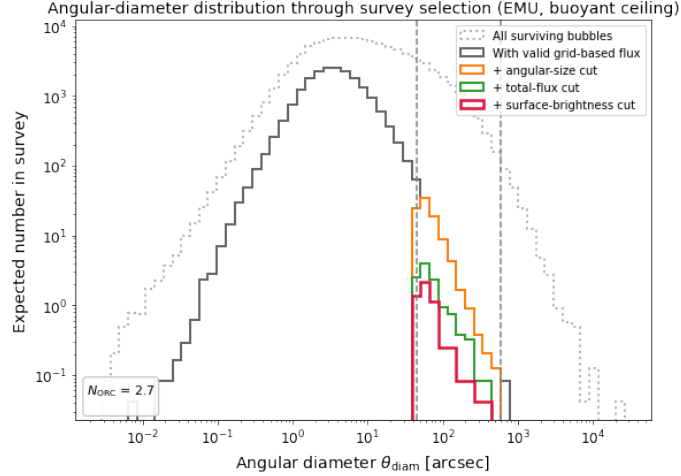}
    \caption{Angular-diameter distribution of the ring population after each ASKAP-EMU survey selection. The dotted grey curve is all surviving bubbles at the time of observation. The solid grey curve is those with a valid grid-based flux, which means the structure of the ring still appears. The orange, green, and red curves add the angular-size, total-flux, and surface-brightness cuts in turn. The intrinsic distribution peaks near $\theta_{\rm diam}\sim\textcolor{red}{3.30}''$, well below the lower cut, so the angular-size selection removes the most of the population and only the large-ring tail survives. The two flux cuts then reduce it further to $N_{\rm ORC}=\textcolor{red}{2.7}$.}
    \label{fig:survey_selection}
\end{figure*}

Figure~\ref{fig:survey_selection} shows the angular-diameter distribution at each selection stage. Most rings extend only a few arcseconds, far below the $45''$ lower cut, so the majority are removed by the angular-size selection alone. The surviving large-ring tail is then further reduced by the total-flux and surface-brightness cuts, leaving $N_{\rm detectable}=\textcolor{red}{5.40}$ and, after the orientation factor, $N_{\rm ORC}=\textcolor{red}{2.7}$.

\section{Discussion}
\label{sec:discussion}
\subsection{\textcolor{black}{Can Shock-Bubble interactions explain the observed ORC abundance?}}
Table~\ref{tab:pipeline_summary} follows the fiducial run from the radio-AGN selection to the expected number of observable ORCs in the ASKAP-EMU Pilot survey. Of the $\sim\!\textcolor{red}{5.71\times10^{7}}$ bubbles inflated across all merger histories, $\textcolor{red}{2.33\times10^{6}}$ remain the bubble structure when their shock arrives and by weight get $N_{\rm all}\approx\textcolor{red}{9.60\times10^{4}}$ expected sources. The following cuts are applied to reduce to the final expected ORC number.

The first is the coverage of the flux grid. A bubble is assigned a flux only if its shock Mach number and ring age fall within the range spanned by the MHD simulations (Table~\ref{tab:mach_age_range}). We excluded the grids outside this range because no ring structures are shown there.  $\textcolor{red}{20.61\%}$ of the surviving bubbles satisfy this, and the rest are set to zero flux.

The total-flux cut reduces $N_\theta\simeq\textcolor{red}{93.41}$ to $N_S\simeq\textcolor{red}{11.50}$, and the surface-brightness cut further to $N_{\rm detectable}\simeq\textcolor{red}{5.40}$. The viewing-angle factor $P_{\rm orientation}=0.5$ then gives
\begin{equation}
N_{\rm ORC}\simeq=\textcolor{red}{2.7^{+3.9}_{-1.6}}.
\end{equation}
\footnote{\textcolor{blue}{The uncertainties are described in the Appendix \ref{sec:uncertainties}.}}
Within the survey volume ($2.24\times10^{8} {\rm Mpc^3}$, $0.01<z<0.55$) and at the ASKAP-EMU Pilot resolution, the model predicts about \textcolor{red}{3} ORCs, against the three found in the Pilot survey \citep{norris_unexpected_2021}. With so few objects, Poisson fluctuations alone allow an observed count anywhere from about one to six for this expectation, and the model uncertainties discussed in Section~\ref{sec:limitations} are larger still. The close numerical agreement is therefore partly coincidental and should not be read as a fit. The model shows that the Shock-Bubble mechanism produces ORCs at the observed order of magnitude.

\begin{figure*}[t]
    \centering
    \includegraphics[width=\textwidth]{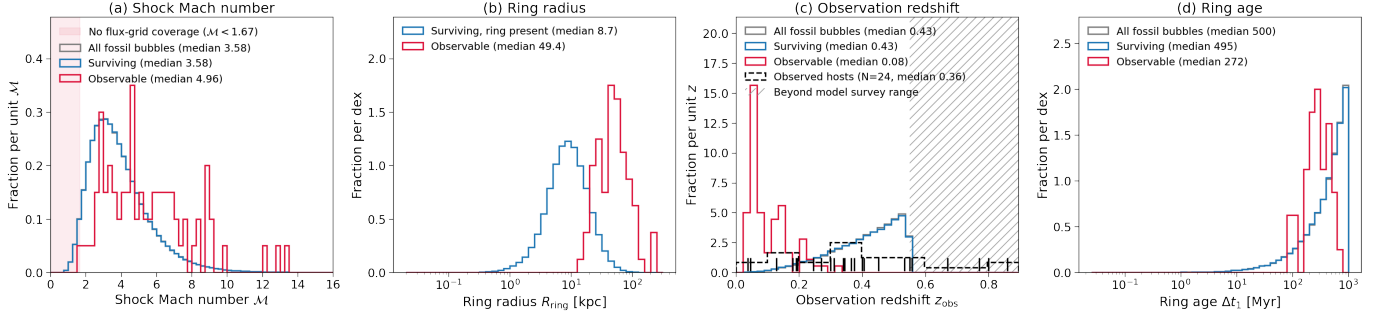}
    \caption{Normalized distributions of (a) the shock Mach number $\mathcal{M}$ (probability density per unit $\mathcal{M}$), (b) the ring radius $R_{\rm ring}$ after the shock (per dex),(c) the observation redshift $z_{\rm obs}$ (per unit $z$), and (d) the ring age $\Delta t_1$, the time since the most recent shock (per dex), for all fossil bubbles (grey), the surviving bubbles (blue), and the observable rings that pass all EMU cuts (red). In panels (a), (c) and (d) the grey and blue curves coincide, because $\mathcal{M}$, $z_{\rm obs}$ and $\Delta t_1$ do not enter the survival criterion. The shaded band in panel (a) marks $\mathcal{M}<1.67$, below the lowest Mach number in the flux grid. In panel (b) only bubbles that currently show a ring have a ring radius, so the blue curve contains the surviving bubbles whose rescaled ring age lies within the simulated ring lifetime. In panel (c) the dashed histogram and the ticks show the redshifts of the galaxies at the centers of 24 known ORCs and candidates, and the hatched region marks $z>0.55$, beyond the redshift range of the model. The sharp edge in panel (d) is the 1 Gyr limit on $\Delta t_1$ in the merger sampling. Median values are given in the legends.}
    \label{fig:ring_mach_radius_redshift}
\end{figure*}

Figure~\ref{fig:ring_mach_radius_redshift} shows the properties of the rings themselves. Of the \textcolor{red}{2,327,873} surviving bubbles, \textcolor{red}{479,767 (20.61\%)} show a ring at the time of observation. The others were either hit by a shock too weak to be covered by the flux grid, or are observed outside the age window over which the simulations predict a ring. The Mach numbers of the surviving bubbles follow the sampled log-normal (median $\mathcal{M}=\textcolor{red}{3.58}$), while the detectable rings come from stronger shocks (median $\mathcal{M}=\textcolor{red}{5.57}$). The detectable rings have a median age of \textcolor{red}{268} Myr since the shock. Their age distribution falls off well before the 1 Gyr limit on $\Delta t_1$, so the limit does not affect the prediction.

Rings are generally small, with a median radius of \textcolor{red}{8.66} kpc. The detectable rings have a median radius of \textcolor{red}{49.90} kpc (diameter \textcolor{red}{99.80} kpc) and a median redshift of $z_{\rm obs}=\textcolor{red}{0.09}$, compared with \textcolor{red}{0.43} for the full population. Both shifts are set by the angular-size cut, which selects rings by $\theta_{\rm diam}=2R_{\rm ring}/D_A(z)$. The $45''$-$600''$ window corresponds to diameters of 18-233 kpc at $z=0.02$, 65-872 kpc at $z=0.08$, and 283-3779 kpc at $z=0.55$. 

The model reproduces the angular sizes of the observed ORCs, $\theta_{\rm diam}\sim1'$, but not their redshift distribution. The galaxies at the centers of the known ORCs have a median redshift of 0.2-0.6, only 2 of the 24 lie below $z=0.1$, and six lie beyond $z=0.55$, outside the range of the model (Fig.~\ref{fig:ring_mach_radius_redshift}c). At $z\sim0.2$-0.6, $1'$ corresponds to ring radii of 100-190 kpc, two to four times larger than the detectable rings in the model. We discuss three possible interpretations of this mismatch:
\begin{enumerate}
\item{The SBI model may produce only part of the ORC population, generating more nearby and smaller rings, while the more distant ORCs are formed in another way. The uncertainties in the predicted incidence are large enough to allow for additional model classes to contribute to the observed numbers.}
\item{The detectable rings in our model are smaller than those implied by the claimed host redshifts because they are at closer redshifts in our population synthesis model. Explaining an ORC population that matches the claimed host galaxy red shift distribution would require larger rings and brighter rings. Larger rings would require more energy in each bubble from more powerful jets (slope b and $\lambda_{\rm jet}$ range of the jet power normalization) and feeding the same bubble for longer time (slower separation $v_{\rm bouy}$). Brighter rings would require a stronger magnetic field in the bubble (smaller $\beta$) or stronger shocks.}
\item{The association with the galaxy at the ring center may be incorrect in at least some ORCs. In the SBI model the ring forms where the shock reaches the fossil bubble, which has moved away from the galaxy that inflated it \citep{wang_odd_2026}, and elliptical ORCs can likewise have offset hosts \citep{shabala_are_2024}. If the observed rings are nearby systems seen in projection against more distant galaxies, their physical sizes and redshifts are overestimated, and the model and the observations can be consistent.}
\end{enumerate}

\begin{table}
    \centering
    \caption{Summary of the fiducial pipeline run ($N_{\rm sample}=10^{6}$ galaxies, $N_\xi=10$, displaced bubbles with $f_{\rm survive}=3$). In the lower block, the percentage after each expected number is the fraction retained relative to the preceding row.}
    \label{tab:pipeline_summary}
    \begin{tabular}{lll}
        \toprule
        Quantity & Value & Description \\
        \hline
        $n_{\rm gal}$ & $1.298\times10^{-2}\ {\rm Mpc^{-3}}$ & Parent galaxy comoving density (Bernardi SMF) \\
        $N_{\rm AGN}$ & $141{,}866$ & Number of selected radio-AGN hosts \\
        $n_{\rm radioAGN}$ & $1.841\times10^{-3}\ {\rm Mpc^{-3}}$ & Selected radio-AGN comoving density \\
        $\log_{10}(M_{\rm h}/M_\odot)$ & $13.11$ (median) & Assigned host halo mass \\
        $V_{\rm survey}$ & $2.239\times10^{8}\ {\rm Mpc^{3}}$ & Effective survey volume ($270\ {\rm deg}^2$, $0.01<z<0.55$) \\
        $f_{\rm recent}$ & $0.0981$ & Fraction of halo draws with $\Delta t_1<1\ {\rm Gyr}$ \\
        $N_{\rm bub}$ & $\textcolor{red}{57{,}068{,}792}$ & Total bubbles after episode grouping \\
        $N_{\rm surv}$ & $\textcolor{red}{2{,}327{,}873}$ & Bubbles surviving until the shock \\
        $n_{\rm recentAGN}$ & $1.806\times10^{-4}\ {\rm Mpc^{-3}}$ & Recent-shock radio-AGN comoving density \\
        $z_{\rm obs}$ & $0.440$ (median) & Per-halo observation redshift, $0.014$-$0.550$ \\
        \hline
        $N_{\rm all}$ & $\textcolor{red}{9.60\times10^{4}}$ & Expected number before any survey cut \\
        $N_{\rm valid}$ & $\textcolor{red}{1.98\times10^{4}}$ ($\textcolor{red}{20.61\%}$) & With valid grid-based flux ($\textcolor{red}{479{,}767/2{,}327{,}873}$) \\
        $N_\theta$ & $\textcolor{red}{93.41}$ ($\textcolor{red}{0.47\%}$) & + angular-size selection \\
        $N_S$ & $\textcolor{red}{11.50}$ ($\textcolor{red}{12.31\%}$) & + total-flux selection \\
        $N_{\rm detectable}$ & $\textcolor{red}{5.40}$ ($\textcolor{red}{46.96\%}$) & + surface-brightness selection \\
        $N_{\rm ORC}$ & $\textcolor{red}{2.7}$ ($50\%$) & $\times P_{\rm orientation}=0.5$ \\
        \hline
    \end{tabular}
\end{table}

\begin{table}
    \centering
    \caption{Ring-age range spanned by each simulated Mach number in the flux grid.}
    \label{tab:mach_age_range}
    \begin{tabular}{lcc}
        \toprule
        Mach number & Min age (Myr) & Max age (Myr) \\
        \hline
        $2$    & $25.70$ & $617.40$ \\
        $2.83$ & $9.10$  & $545.50$  \\
        $4$    & $12.90$ & $565.90$  \\
        $5.66$ & $9.10$  & $363.80$  \\
        $8$    & $6.40$  & $205.80$  \\
        $16$   & $1.6$  & $96.50$   \\
        \hline
    \end{tabular}
\end{table}

\subsection{\textcolor{black}{What controls the incidence?}}
\label{sec:what_controls}
The predicted incidence is controlled by two kinds of input: the survey, which is the strongest factor, and a small number of internal model choices. The same bubble population yields $N_{\rm ORC}\simeq=\textcolor{red}{2.7^{+3.9}_{-1.6}}$ for the ASKAP-EMU Pilot but $\simeq\textcolor{red}{268^{+331}_{-148}}$ for LoTSS (Table~\ref{tab:threeruns}); the difference is set entirely by the observing survey, and we return to it in Section~\ref{sec:lotss}. Here we consider what sets the number for a fixed survey.

Two timescales determine whether a bubble contributes to the ring population. The first is the \emph{inflation time}, $t_{\rm on}$, over which the jet injects energy and grows the bubble to $R_{\rm bub}$. It is drawn from the AGN episode-lifetime distribution with mean $\langle t_{\rm jet}\rangle=45 {\rm Myr}$ and, together with the jet power, follows from the radio AGN incidence and Eddington-ratio distributions adopted in Section~\ref{sec:bubble_energies}. The one assumption at this stage is that hosts undergo recurrent AGN episodes, so that the fossil bubble a shock encounters later is built by several inflation events rather than a single burst. This shifts the bubble energy and radius distributions toward larger, more energetic bubbles.

The second is the \emph{survival time}, over which the inflated bubble must remain coherent until the shock arrives. We set $t_{\rm survive}=\min( R_{\rm bub}/v_{\rm buoy},  t_{\rm survive,max})$ (Eq.~\ref{eq:tsurvive}). The buoyant rise time $R_{\rm bub}/v_{\rm buoy}$, with $v_{\rm buoy}=f_{\rm buoy}c_s$, is the time for the bubble to rise about its own radius through the halo atmosphere. We adopt $f_{\rm buoy}=1$, so that bubbles rise at approximately the ambient sound speed, as expected for buoyant radio bubbles \citep{churazov_evolution_2001}, and the term reduces to the sound-crossing time $R_{\rm bub}/c_s$. The ceiling $t_{\rm survive,max}=10 {\rm Gyr}$ is physical, imposed because a bubble cannot last beyond the available cosmic time. The result is nonetheless robust to this choice because the surviving population is governed by the physical timescale $R_{\rm bub}/c_s$.

Bubbles that survive until the shock arrives have a median $t_{\rm on}\simeq\textcolor{red}{55} {\rm Myr}$ (16th-84th percentile range \textcolor{red}{10-245} Myr), comparable to the input mean $\langle t_{\rm jet}\rangle=45 {\rm Myr}$. Survival therefore depends on the injection-time distribution only weakly. This raises the question of how the predictions depend on the duty cycle. Changing $\langle t_{\rm duty}\rangle$ from 4.5 to 45 and 90 Myr changes the number of surviving bubbles from \textcolor{red}{4,217,625} to \textcolor{red}{2,327,873} and \textcolor{red}{1,747,253}, because shorter episodes divide the jet energy among more, smaller bubbles. The number of ORCs changes much less than the duty cycle: by a factor of \textcolor{red}{1.46} for EMU (\textcolor{red}{4.66} to \textcolor{red}{3.20}) and \textcolor{red}{1.65} for LoTSS (\textcolor{red}{351.67} to \textcolor{red}{212.84}) over a factor of 20 in $\langle t_{\rm duty}\rangle$. The properties of the rings change in a consistent way. Their median radius \textcolor{red}{changes little} (from \textcolor{red}{47.50} to \textcolor{red}{42.50} kpc for EMU, and from \textcolor{red}{62.10} to \textcolor{red}{65.50} kpc for LoTSS), and in LoTSS, where the sample is large enough, the median redshift \textcolor{red}{is} \textcolor{red}{0.128}, \textcolor{red}{0.139} and \textcolor{red}{0.132}. Longer duty cycles produce fewer\textcolor{red}{ rings}. Over the range we tested, however, the shift in redshift is small compared with the offset from the observed hosts (Section~\ref{sec:redshift_offset}), and it cannot by itself bring the predicted redshifts up to those of the known ORCs.

Based on this discussion, the predictions by the model are surprisingly robust against the assumed duty cycle, and we conclude that ORC incidence is not a sensitive probe of this parameter.  Conversely, we find that the match of the predicted incidence with the observed numbers is an independent supporting evidence that the AGN kinetic power functions of \citet{igo_global_2025} and \citet{oshea_shooting_2025} are realistic representations of the underlying ANG jet population..

\subsection{Expected host environments}
\begin{figure}
    \centering
    \includegraphics[width=0.5\columnwidth]{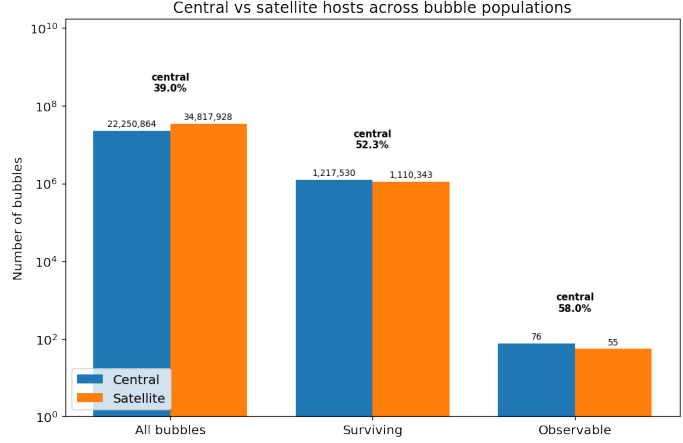}
    \caption{Central versus satellite hosts across the bubble populations in the EMU configuration: all bubbles, those surviving the coherence cut, and the observable ORCs. The central fraction rises through the selection, from $\textcolor{red}{39.00\%}$ of all bubbles to $\textcolor{red}{58.00\%}$ of observable ORCs, so detectable ORCs are predominantly hosted by central galaxies even though satellites are more numerous overall.}
    \label{fig:central_satellite}
\end{figure}

The detectable ORCs are hosted by halos with a median mass of $\log_{10}(M_{\rm h}/M_\odot)\simeq\textcolor{red}{13.36}$ for EMU and \textcolor{red}{13.37} for LoTSS (Figure~\ref{fig:halomass_joint}), i.e.\ massive galaxies and small groups. Most are hosted by central galaxies. Central galaxies account for \textcolor{red}{39.00\%} of all bubbles and \textcolor{red}{52.30\%} of the survivors, but for \textcolor{red}{58.00\%} the detectable ORCs in EMU and \textcolor{red}{62.50\%} in LoTSS (Figure~\ref{fig:central_satellite}). Satellites move through the halo faster than central galaxies ($f_{v{\rm AGN}}=1$ compared with 0.3), so they leave their bubble sooner and divide the jet energy among more, less energetic bubbles. 

The model therefore makes a prediction for where ORCs should be found. If they form through Shock-Bubble interactions, they should mostly be associated with central galaxies in halos of $\sim10^{13}\,M_\odot$, where the ambient gas density is $n_{\rm e}\approx 10^{-4}\,{\rm cm^{-3}}$, and with halos that have had a major merger within the last Gyr. The ring itself need not be centered on its host. It forms where the shock reaches the fossil bubble, which has moved away from the galaxy that inflated it. ORCs could therefore serve as probes of their environment. 

\subsection{Predictions for LoTSS}\label{sec:lotss}
We ran the same AGN bubble and merger shock population through the selection for LOFAR Two-metre Sky Survey (LoTSS), changing only the observing parameters and cuts: an observing frequency of 144 MHz, a $6''$ beam, an image noise of $\sigma_{\rm rms}\simeq89\ \mu{\rm Jy beam^{-1}}$, and an angular-diameter range of $40''$-$300''$ \citep{gasperin_single-_2026}. For the effective search area we take the LoTSS DR3 footprint (19,035 deg$^2$)\citep{shimwell_lofar_2026} reduced by the two cuts \citet{gasperin_single-_2026} applied in their search: a Galactic-latitude cut $|b|>20^\circ$, which keeps a fraction $1-\sin20^\circ\simeq0.66$ of the sky, and a noise cut of $0.1\ {\rm mJy beam^{-1}}$, which we estimate to keep $\simeq0.6$ of the footprint given the DR3 median noise of $92\ \mu{\rm Jy beam^{-1}}$. This gives an effective area of $\sim7.5\times10^{3}\ {\rm deg^2}$. Over $0.01<z<0.55$, the predicted number of ORCs detectable by LoTSS is
\begin{equation}
N_{\rm ORC}^{\rm LoTSS}=\textcolor{red}{268^{+331}_{-148}}.
\end{equation}

Table~\ref{tab:threeruns} summarizes the two survey configurations. Both draw on the identical fossil-bubble and shock population, and differ only in the observing survey: EMU at 944 MHz over 270 deg$^2$, and LoTSS at 144 MHz over an effective $\sim7.5\times10^{3}$ deg$^2$. The difference reflects the survey rather than any change in the underlying population.

Two effects make the LoTSS number much larger than the EMU one: First, the effective survey volume for LoTSS is about \textcolor{red}{27.83} times larger ($\textcolor{red}{6.23\times10^{9}}\ {\rm Mpc^3}$). And second, the aged rings are much brighter at low frequency because a ring that has cooled below detectability at 944 MHz can still carry substantial flux density at 144 MHz, so LoTSS recovers rings that have already faded from EMU. 

Most of these $\sim\!\textcolor{red}{268}$ rings are steep-spectrum and of low surface brightness, and would be hard to pick out. 12 of the 18 are detected to have high frequency counterpart at 1.4 GHz. We extrapolate each ring's surface brightness from 144 MHz to 1.4 GHz using its own two-point spectral index,
\begin{equation}
S_{\rm beam}(1.4 {\rm GHz}) = S_{\rm beam}(144 {\rm MHz})
\left(\frac{1400}{144}\right)^{\alpha_{144}^{1400}},
\end{equation}
and re-apply the surface-brightness threshold. The predicted population has a steep median two-point index, $\alpha_{144}^{1400}\simeq\textcolor{red}{-2.12}$. Only $\simeq\textcolor{red}{2.78\%}$ of the population is now observable at 1.4 GHz, giving
\begin{equation}
N_{\rm ORC}^{\rm LoTSS, 1.4 GHz}\simeq \textcolor{red}{7.46^{+9.51}_{-4.18}}
\end{equation}

The counterpart-bright subset ($\lesssim\textcolor{red}{7.46}$) is smaller   than the 12 sources with $1.4 {\rm GHz}$ detections, while the full $144 {\rm MHz}$ population is one to two orders of magnitude larger. On the other hand, the reported sample is not complete. Specifically, in \citet{gasperin_single-_2026} it was \emph{''assembled by visual inspection of a filtered subset of the survey and note that some candidates are missed, so a prediction somewhat above the number recovered by an incomplete search is expected''}. 

The observed numbers in the two surveys give a further constraint that does not depend on the model. Scaling the three EMU Pilot ORCs by area alone gives $\sim80$ ORCs expected in LoTSS, whereas 18 have been identified. The model predicts a still larger number simply by area scaling alone. It is possible that many ORCs in LoTSS DR3 have not yet been identified. The missing rings are most likely the faint, steep-spectrum systems that the model places at low redshift. 

\begin{table*}[t]
\centering
\caption{The two survey configurations for the fiducial model. Both use the same fossil-bubble population and differ only in the survey. The valid-flux fraction is the fraction of surviving bubbles with a nonzero grid flux at the observing frequency. $N_{\rm detectable}$ is the expected number passing all survey cuts, and $N_{\rm ORC}=P_{\rm orient}N_{\rm detectable}$ with $P_{\rm orient}=0.5$.}
\label{tab:threeruns}
\begin{tabular}{lll}
\hline\hline
 & EMU & LoTSS \\
\hline
Observing frequency                       & 944\,MHz & 144\,MHz \\
Survey area                               & 270\,deg$^2$ & $7.5\times10^{3}$\,deg$^2$ \\
Redshift range                            & $0.01$-$0.55$ & $0.01$-$0.55$ \\
$V_{\rm survey}$                          & $2.24\times10^{8}$\,Mpc$^3$ & $6.23\times10^{9}$\,Mpc$^3$ \\
Beam $\theta_{\rm beam}$                  & $11.96''$ & $6''$ \\
$\sigma_{\rm rms}$                        & $25\,\mu$Jy\,beam$^{-1}$ & $89\,\mu$Jy\,beam$^{-1}$ \\
Angular-diameter range                    & $45''$-$600''$ & $40''$-$300''$ \\
Valid-flux fraction                       & $\textcolor{red}{20.61\%}$ & $\textcolor{red}{40.16\%}$ \\
$N_{\rm detectable}$                      & $\textcolor{red}{5.40}$ & $\textcolor{red}{535.83}$ \\
$N_{\rm ORC}$                             & $\textcolor{red}{2.7}$ & $\textcolor{red}{268}$ \\
\hline
\end{tabular}
\end{table*}

\subsection{Caveats}
Given the chain of Monte Carlo steps required, our model rests on a large number of previously established relations and assumptions. Below, we list the adopted parameters or simplified assumptions made and discuss whether and to what extent uncertainties in those assumption affect the reliability ouf our predictions.
\label{sec:limitations}\label{sec:caveats}

\begin{enumerate}

\item \textit{Radio-AGN incidence extrapolation and no redshift evolution.} The incidence plane of \citet{oshea_shooting_2025} rises with stellar mass and formally exceeds unity above $\log_{10}M_\star\approx10.92$, where we cap $P_{\rm AGN}=1$ (Equation~\eqref{eq:pagn_cap}). The incidence relation, the stellar mass function, and the halo statistics are measured locally ($z\lesssim0.4$) and applied without evolution over the full survey range, $0.01<z<0.55$. The kinetic-Eddington-ratio range $-6\leq\log_{10}\lambda_{\rm jet}\leq-1$ is also a local, bounded choice. The predicted detectable rings lie at low redshift (median $z_{\rm obs}=\textcolor{red}{0.09}$ for EMU and \textcolor{red}{0.14} for LoTSS), where these relations apply. The known ORCs were associated with their central galaxies at $z\sim0.2$-0.6, where evolution matters more. Over this range the halo merger rate per unit time is higher by a factor $(1+z)E(z)\approx1.3$-2.1, which would raise the number of rings at these redshifts, while the group gas is denser by $E(z)^2\approx1.2$-1.7, which makes bubbles smaller. Including this evolution would therefore increase the predicted number of ORCs at $z>0.2$, but it would not produce the larger rings needed to pass the angular-size cut at these redshifts, so it does not remove the redshift offset.

\item \textit{Halo evolution and redshift.} The descendant halo mass $M_0$ is held fixed over its history (no growth), and the CSMF and halo mass function are evaluated at $z=0$, while the halo structure that sets $P_{\rm env}$ is evaluated at $z_{\rm first}$. Neglecting halo growth has little effect on the recent-merger statistics that drive the calculation, because the merger rate depends only weakly on mass ($\propto M_0^{0.133}$) \citep{fakhouri_merger_2010}. Evaluating the profile at $z_{\rm first}$ fixes $P_{\rm env}$, and hence the bubble radius and flux, but their change should be small within our redshift range. Letting the halo grow would place earlier bubbles in smaller progenitors and change $P_{\rm env}$ along the history. Higher pressure shrinks the bubbles, as $R_{\rm bub}\propto\rho^{-1/5}$ in the wind phase and $\propto P_{\rm env}^{-1/3}$ in pressure balance, but makes the rings brighter. So the net shift in detectable count should be small.

\item \textit{Halo gas and galaxy position.} The $\beta$-model uses $\beta_{\rm gas}=2/3$ and core fraction $f_{\rm core}=0.1$. All satellite hosts are placed following the profile of \citet{budzynski_radial_2012}, drawn out to $2R_{200{\rm m}}$. Because that profile was measured only within $\sim R_{200{\rm m}}$, placing satellites at larger radii extrapolates it beyond its calibrated range. Satellites make up \textcolor{red}{61.00\%} of all bubbles but only \textcolor{red}{42.00\%} of the detectable ORCs for EMU and \textcolor{red}{37.50\%} for LoTSS. The outermost satellites sit in lower-density gas, so at fixed energy their bubbles are larger by $(P_{\rm core}/P_{\rm sat})^{1/3}\approx3$, and since $t_{\rm survive}\propto R_{\rm bub}$ they also survive longer. Two effects decrease the satellite fraction moderately. First, satellites move through the halo faster than centrals, so they leave their bubble sooner and spread the jet energy over more, less energetic bubbles. Second, low pressure makes a ring faint. Most satellite rings fail the flux cuts even when they are large enough to be resolved.

\item \textit{Particle acceleration.} We assume that the shock passage re-accelerates the previously existing population of non-thermal particles and start the spectral evolution calculation with a freshly injected powerlaw distribution, as described in \citet{wang_odd_2026}. This is clearly a simplification and presents one of the key factors of uncertainty in the overall flux at high frequencies and therefore also in the predicted number of ORCs. A detailed treatment of particle re-acceleration at shocks propagating through pre-existing non-thermal plasma is beyond the scope of this paper.

\item \textit{AGN episode lifetimes.} The mean episode lifetime is fixed to one Salpeter time, $\langle t_{\rm jet}\rangle=45\ {\rm Myr}$, with a log-normal distribution whose standard deviation equals its mean $\sigma(t_{\rm jet})/\langle t_{\rm jet}\rangle=1$. We also assume the AGN is active continuously with different powers and durations between mergers with no gaps. This is reasonable because the lower incidence of radio AGN at low stellar mass mainly reflects galaxies that are never radio AGN, which our selection excludes, while the more massive, early-type galaxies are all radio AGN at some level.  Because an AGN spends far more time at low than at high jet power, periods of low activity are already represented by episodes with small $\lambda_{\rm jet}$. 
The detectable bubbles have a median total jet-on time of \textcolor{red}{126} Myr for EMU and \textcolor{red}{143} Myr for LoTSS, accumulated over a median of \textcolor{red}{3} episodes. Detectable ORCs therefore come from the bubbles that were inflated over several episodes and grew large. Changing $\langle t_{\rm duty}\rangle$ affects only weakly their number (Section~\ref{sec:what_controls} shown in Appendix~\ref{app:sensitivity}). From 4.5 to 90 Myr, $N_{\rm ORC}$ falls by a factor of \textcolor{red}{1.46} for EMU (\textcolor{red}{4.66} to \textcolor{red}{3.20}) and \textcolor{red}{1.65} for LoTSS (\textcolor{red}{351.67} to \textcolor{red}{212.84}), while the median ring radius \textcolor{red}{changes from 47.50 to 42.50 kpc for EMU} and, in LoTSS, the median redshift \textcolor{red}{from} \textcolor{red}{0.128} to \textcolor{red}{0.132}. Longer duty cycles thus produce fewer\textcolor{red}{ rings}, and shorter duty cycles more\textcolor{red}{ rings}. Because the number stays within the same order of magnitude, the observed counts alone cannot constrain the duty cycle.

\item \textit{AGN double lobes.} Each episode is modeled as a single bubble carrying the full injected energy, whereas real AGN jets are bipolar and could inflate two lobes. Splitting the energy between two lobes would give two smaller rings. This simplification affects the ring-size distribution and is directly relevant to the double-headed ORCs (four of the eighteen LoTSS sources; \citealt{gasperin_single-_2026}), which the single-bubble model does not reproduce.

\item \textit{Black-hole mass normalization.} We adopt $M_{\rm BH}=0.002 M_\star$, following \citet{marconi_relation_2003}, the same mass-independent ratio used by \citet{igo_global_2025} on whose radio-AGN incidence framework we build. This keeps our jet-power normalization consistent with theirs. We apply it as a single fixed ratio with no intrinsic scatter. Varying the ratio over the plausible range $0.002$-$0.005$ has little impact on $N_{\rm ORC}$.

\item \textit{Bubble growth and survival.} The buoyant-rise velocity is $v_{\rm buoy}=f_{\rm buoy}c_s$ with $f_{\rm buoy}=1$, so bubbles rise at the ambient sound speed, which is at the fast end of the $\sim0.5$-$1 c_s$ range inferred by \citet{churazov_evolution_2001}, and thus the short survival time makes the count conservative. We assume that satellites move through the intragroup medium at roughly the velocity dispersion ($f_{v{\rm AGN}}=1$), while central galaxies move much more slowly relative to the halo ($f_{v{\rm AGN,cen}}=0.3$). A slower rise or smaller drift would increase bubble survival times, and also keep the host inside its bubble longer, so that more episodes add to the same bubble and produce larger, more energetic bubbles, which would raise $N_{\rm ORC}$.

\item \textit{Bubble plasma $\beta$.} The bubble plasma $\beta$ is fixed at $10$ for all bubbles. A higher $\beta$ means a weaker field for the same gas pressure, and therefore fainter rings. Raising it to 20 lowers $N_{\rm ORC}$ from \textcolor{red}{2.7} to \textcolor{red}{1.57} for EMU and from \textcolor{red}{268} to \textcolor{red}{170.96} for LoTSS. Lowering it to 1 raises them to \textcolor{red}{5.03} and \textcolor{red}{371.75}, respectively (Appendix~\ref{app:sensitivity}). All three values sit within an order of magnitude of the observed number from the fiducial run, so the $\beta$ choice does not change our conclusion, which is also consistent with the results of simulation analysis in \citet{wang_odd_2026}.

\item \textit{Flux-grid coverage.} Only $\sim\textcolor{red}{20.61\%}$ of surviving bubbles fall within the Mach-number ($1.67\leq\mathcal{M}\leq16$) and ring-age coverage of the MHD grid (Table~\ref{tab:mach_age_range}). Outside the grid, the shocked bubble does not have the edge-brightened, ring-like morphology that defines an ORC. Shocks too weak to compress the bubble into a coherent ring ($\mathcal{M}<1.67$), and evolutionary phases before the ring has formed or after it has dispersed into a diffuse or disrupted structure, do not produce ORC-like sources. Excluding these bubbles is therefore the correct treatment.

\item \textit{Ring sizes and redshifts.} The model reproduces the angular sizes of the observed ORCs but places them at closer redshifts. The detectable rings are physically small, with median radii of \textcolor{red}{49.90} kpc for EMU and \textcolor{red}{65.00} kpc for LoTSS, at a median $z_{\rm obs}$ of \textcolor{red}{0.09} and \textcolor{red}{0.14}. The known ORCs have central galaxies at $z\sim 0.2$-0.6, which implies radii of 100-200 kpc, matching their reported sizes. Bubbles could reach these sizes if they expand into lower-pressure regions away from the host before the shock arrives, since we evaluate $P_{\rm env}$ at the galaxy host position. Alternatively, some observed rings may be nearby systems seen in projection in the background galaxies, in which case their redshifts should not be associated with the central galaxy redshifts and their physical sizes are overestimated.
\label{sec:redshift_offset}
\end{enumerate}

\section{Conclusions}
\label{sec:concl}
We built a Monte Carlo population-synthesis model to test whether Shock-Bubble interactions can account for the observed number of odd radio circles. The model draws galaxies from the stellar mass function and assigns radio-AGN activity and host halo masses from local scaling relations. It then follows the merger histories of the host halos, inflates AGN bubbles, and tracks whether each bubble survives until a merger shock reaches it. Synthetic radio fluxes and ring geometries are calculated from a grid of magnetohydrodynamic Shock-Bubble simulations covering Mach numbers 1.67-16 \citep{wang_odd_2026}. We applied the selection of each survey to this population and compared the predicted numbers with the known ORCs.

For the EMU Pilot Survey at 944 MHz the fiducial model predicts \textcolor{red}{2.7} detectable ORCs, consistent with the three found in the Pilot \citep{norris_unexpected_2021}. For LoTSS at 144 MHz it predicts \textcolor{red}{268}, of which \textcolor{red}{7.46} would also be detected at 1.4 GHz, compared with 12 observed. Both surveys are reproduced by the same fossil-bubble population, only changing the observing frequency, resolution, noise and survey area.

The predicted hosts are mostly central galaxies in group-scale halos, with a median $\log_{10}(M_{\rm h}/M_\odot)\simeq\textcolor{red}{13.36}$ for EMU and \textcolor{red}{13.37} for LoTSS. The central galaxy fraction increases after the survey selections from \textcolor{red}{39.00\%} of all bubbles to \textcolor{red}{58.00\%} (EMU) and \textcolor{red}{62.50\%} (LoTSS) of the detectable ORCs. The rings that pass the cuts are large and bright, and they come mostly from powerful jets in massive central galaxies.

The model reproduces the angular sizes of the observed ORCs but places them at lower redshift than those inferred from the putative central galaxies of the observed ORCs. The detectable rings have median radii of \textcolor{red}{49.90} kpc (EMU) and \textcolor{red}{65.00} kpc (LoTSS) and median redshifts of \textcolor{red}{0.09} and \textcolor{red}{0.14}. The known ORCs have central galaxies at $z\sim0.2$-0.6, which would imply radii of 100-190 kpc. Either some observed rings are associated with the wrong host, or the model produces too few large bubbles detectable as ORCs at larger redshifts.

Most of the predicted LoTSS population consists of steep-spectrum, low-surface-brightness rings that are visible only at low frequency. Of the detectable rings at 144 MHz, \textcolor{red}{40.00\%} have no emission left at 1.4 GHz in the model, and the rest have a median two-point spectral index of $\textcolor{red}{-2.12}$ between 144 MHz and 1.4 GHz. The model therefore predicts a large population of ORCs only at low frequency, and they should already exist in LoTSS DR3. Either the model overpredicts the low-frequency population, or the search has missed many of these rings. A future data search for diffusive rings at low frequencies with no high-frequency counterparts could answer this.

These results are robust to the model parameters we tested. Varying the bubble plasma $\beta$ from 1 to 20 changes the EMU prediction from \textcolor{red}{5.03} to \textcolor{red}{1.57}, and varying the mean jet-episode duration from 4.5 to 90 Myr changes it from \textcolor{red}{4.66} to \textcolor{red}{3.20}. In every case the prediction stays within a factor of two of the observed number. The two steps that remove the most bubbles are the formation of a ring and the angular-size cut. Only \textcolor{red}{20.61\%} of the surviving bubbles show a ring at the time of observation, because the rest were hit by shocks too weak for the simulation grid or are seen outside the simulated ring lifetime. The angular-size cut then removes most of the remaining rings, which are too small to be resolved.

We conclude that Shock-Bubble interactions can produce ORCs at the observed rate and in plausible host environments. The main limitations of the model are the assumption of a single bubble per AGN episode, which cannot produce double-headed ORCs, and the small physical sizes of the predicted rings compared with those inferred for the known ORCs. The model can be applied directly to upcoming wide-field and low-frequency surveys, which will test whether the predicted population of faint, steep-spectrum rings exists.

\section*{Acknowledgments}

We thank Larry Rudnick, Ludwig B\"{o}ss, Marcur Br\"{u}ggen, Rachel Somerville, and Talia O'Shea for helpful discussions that aided in the preparation of this manuscript. AT acknowledges support by NASA
80NSSC26K0343, 
80NSSC22K0031, 
80NSSC22K0799, 
80NSSC18K0565 
and 80NSSC21K1746 
grants, and by the NSF grants 
AST-2009884, 
AST-2107839, 
AST-1815304, 
AST-1911080, 
AST-2206471, 
AST-2407475, 
AST-2510570. 

\section*{AI assistance in code development}
The authors thank Claude Sonnet 5.5 (Anthropic) to help write specific parts of the Python code, such as individual functions, plotting routines and data bookkeeping, following steps defined by the authors. The structure of the population synthesis pipeline, the physical model, the choice of parameters, the selection criteria and the statistical methods were designed by the authors. The authors reviewed and tested all code, checked its outputs against independent calculations, and are responsible for the final code and results.

\bibliographystyle{apsrev4-1}

\bibliography{ORC_population}

\appendix
\section{Sensitivity to model parameters}
\label{app:sensitivity}

We test how the predicted ORC numbers depend on three choices that are not fixed by observations: the plasma $\beta$ of the fossil bubbles, the mean duration of the AGN jet episodes $t_{\rm Sal}$, and the random realization of the Monte Carlo simulation. In each run we change one quantity from the fiducial model ($\beta=10$, $t_{\rm Sal}=45$ Myr, seed 20) and keep everything else fixed. EMU and LoTSS runs with the same seed and $t_{\rm Sal}$ use the same bubble population and differ only in the survey selection. Tables~\ref{tab:sens_emu} and \ref{tab:sens_lotss}, \ref{tab:sens_lotss_1p4}
list the expected numbers after each selection stage for EMU and LoTSS.

\begin{table*}[h]
\centering
\caption{Sensitivity of the EMU predictions at 944 MHz. $N_{\rm ORC}$ is the expected number of ORCs. All other columns are medians over the Monte Carlo rings that pass all survey cuts: bubble energy $E_{\rm bub}$, ring radius $R_{\rm ring}$, observation redshift $z_{\rm obs}$, ambient electron density at the bubble when the shock arrives $n_{\rm e}$, total jet-on time $t_{\rm on}$ with the number of episodes per bubble in parentheses, and shock Mach number $\mathcal{M}$. The fiducial run is in bold.}
\label{tab:sens_emu}
\begin{tabular}{lccccccc}
\hline
Run & $N_{\rm ORC}$ & $E_{\rm bub}$ [erg] & $R_{\rm ring}$ [kpc] & $z_{\rm obs}$ & $n_{\rm e}$ [cm$^{-3}$] & $t_{\rm on}$ [Myr] ($N_{\rm ep}$) & $\mathcal{M}$ \\
\hline
\textbf{EMU-seed20-beta10-t45} & \textbf{\textcolor{red}{2.7}} & \textbf{\textcolor{red}{$1.01\times10^{58}$}} & \textbf{\textcolor{red}{49.90}} & \textbf{\textcolor{red}{0.09}} & \textbf{\textcolor{red}{$1.50\times10^{-4}$}} & \textbf{\textcolor{red}{126 (3)}} & \textbf{\textcolor{red}{5.57}} \\
EMU-seed20-beta1-t45   & \textcolor{red}{5.03} & \textcolor{red}{$1.16\times10^{58}$} & \textcolor{red}{56.70} & \textcolor{red}{0.09} & \textcolor{red}{$6.82\times10^{-5}$} & \textcolor{red}{144 (3)} & \textcolor{red}{4.43} \\
EMU-seed20-beta20-t45  & \textcolor{red}{1.57} & \textcolor{red}{$1.07\times10^{58}$} & \textcolor{red}{49.20} & \textcolor{red}{0.09} & \textcolor{red}{$1.95\times10^{-4}$} & \textcolor{red}{118 (3)} & \textcolor{red}{6.02} \\
EMU-seed34-beta10-t45  & \textcolor{red}{3.71} & \textcolor{red}{$1.05\times10^{58}$} & \textcolor{red}{49.50} & \textcolor{red}{0.09} & \textcolor{red}{$1.54\times10^{-4}$} & \textcolor{red}{129 (3)} & \textcolor{red}{5.38} \\
EMU-seed20-beta10-t4p5 & \textcolor{red}{4.66} & \textcolor{red}{$9.20\times10^{57}$} & \textcolor{red}{47.50} & \textcolor{red}{0.09} & \textcolor{red}{$1.33\times10^{-4}$} & \textcolor{red}{102 (23)} & \textcolor{red}{5.04} \\
EMU-seed20-beta10-t90  & \textcolor{red}{3.20} & \textcolor{red}{$7.16\times10^{57}$} & \textcolor{red}{42.50} & \textcolor{red}{0.08} & \textcolor{red}{$1.86\times10^{-4}$} & \textcolor{red}{124 (2)} & \textcolor{red}{4.82} \\
\hline
\end{tabular}
\end{table*}

\begin{table*}
\centering
\caption{Sensitivity of the LoTSS predictions at 144 MHz. $N_{\rm ORC}$ is the expected number of ORCs. All other columns are medians over the Monte Carlo rings that pass all survey cuts: bubble energy $E_{\rm bub}$, ring radius $R_{\rm ring}$, observation redshift $z_{\rm obs}$, ambient electron density at the bubble when the shock arrives $n_{\rm e}$, total jet-on time $t_{\rm on}$ with the number of episodes per bubble in parentheses, and shock Mach number $\mathcal{M}$. The fiducial run is in bold.}
\label{tab:sens_lotss}
\begin{tabular}{lccccccc}
\hline
Run & $N_{\rm ORC}$ & $E_{\rm bub}$ [erg] & $R_{\rm ring}$ [kpc] & $z_{\rm obs}$ & $n_{\rm e}$ [cm$^{-3}$] & $t_{\rm on}$ [Myr] ($N_{\rm ep}$) & $\mathcal{M}$ \\
\hline
\textbf{LoTSS-seed20-beta10-t45} & \textbf{\textcolor{red}{268}} & \textbf{\textcolor{red}{$3.87\times10^{58}$}} & \textbf{\textcolor{red}{65.00}} & \textbf{\textcolor{red}{0.14}} & \textbf{\textcolor{red}{$1.71\times10^{-4}$}} & \textbf{\textcolor{red}{143 (3)}} & \textbf{\textcolor{red}{5.07}} \\
LoTSS-seed20-beta1-t45   & \textcolor{red}{371.75} & \textcolor{red}{$5.02\times10^{58}$} & \textcolor{red}{81.40} & \textcolor{red}{0.16} & \textcolor{red}{$9.09\times10^{-5}$} & \textcolor{red}{169 (4)} & \textcolor{red}{4.45} \\
LoTSS-seed20-beta20-t45  & \textcolor{red}{170.96} & \textcolor{red}{$4.97\times10^{58}$} & \textcolor{red}{64.50} & \textcolor{red}{0.13} & \textcolor{red}{$2.08\times10^{-4}$} & \textcolor{red}{140 (3)} & \textcolor{red}{5.50} \\
LoTSS-seed34-beta10-t45  & \textcolor{red}{298.18} & \textcolor{red}{$4.26\times10^{58}$} & \textcolor{red}{69.70} & \textcolor{red}{0.14} & \textcolor{red}{$1.66\times10^{-4}$} & \textcolor{red}{152 (4)} & \textcolor{red}{5.01} \\
LoTSS-seed20-beta10-t4p5 & \textcolor{red}{351.67} & \textcolor{red}{$3.01\times10^{58}$} & \textcolor{red}{62.10} & \textcolor{red}{0.13} & \textcolor{red}{$1.67\times10^{-4}$} & \textcolor{red}{122 (27)} & \textcolor{red}{5.03} \\
LoTSS-seed20-beta10-t90  & \textcolor{red}{212.84} & \textcolor{red}{$4.66\times10^{58}$} & \textcolor{red}{65.50} & \textcolor{red}{0.13} & \textcolor{red}{$1.85\times10^{-4}$} & \textcolor{red}{153 (2)} & \textcolor{red}{4.85} \\
\hline
\end{tabular}
\end{table*}

\begin{table*}
\centering
\caption{As Table~\ref{tab:sens_lotss}, restricted to the LoTSS ORCs that would also be detected at 1.4 GHz. $N_{\rm ORC}^{1.4 {\rm GHz}}$ is their expected number.}
\label{tab:sens_lotss_1p4}
\begin{tabular}{lccccccc}
\hline
Run & $N_{\rm ORC}^{1.4 {\rm GHz}}$ & $E_{\rm bub}$ [erg] & $R_{\rm ring}$ [kpc] & $z_{\rm obs}$ & $n_{\rm e}$ [cm$^{-3}$] & $t_{\rm on}$ [Myr] ($N_{\rm ep}$) & $\mathcal{M}$ \\
\hline
\textbf{LoTSS-seed20-beta10-t45} & \textbf{\textcolor{red}{7.46}} & \textbf{\textcolor{red}{$2.39\times10^{58}$}} & \textbf{\textcolor{red}{44.60}} & \textbf{\textcolor{red}{0.09}} & \textbf{\textcolor{red}{$4.02\times10^{-4}$}} & \textbf{\textcolor{red}{109 (3)}} & \textbf{\textcolor{red}{5.07}} \\
LoTSS-seed20-beta1-t45   & \textcolor{red}{10.33} & \textcolor{red}{$3.54\times10^{58}$} & \textcolor{red}{45.70} & \textcolor{red}{0.09} & \textcolor{red}{$2.85\times10^{-4}$} & \textcolor{red}{124 (3)} & \textcolor{red}{3.62} \\
LoTSS-seed20-beta20-t45  & \textcolor{red}{6.88} & \textcolor{red}{$3.38\times10^{58}$} & \textcolor{red}{41.40} & \textcolor{red}{0.08} & \textcolor{red}{$4.09\times10^{-4}$} & \textcolor{red}{107 (3)} & \textcolor{red}{5.25} \\
LoTSS-seed34-beta10-t45  & \textcolor{red}{11.97} & \textcolor{red}{$3.55\times10^{58}$} & \textcolor{red}{45.60} & \textcolor{red}{0.11} & \textcolor{red}{$3.44\times10^{-4}$} & \textcolor{red}{117 (3)} & \textcolor{red}{4.79} \\
LoTSS-seed20-beta10-t4p5 & \textcolor{red}{20.08} & \textcolor{red}{$6.91\times10^{57}$} & \textcolor{red}{30.30} & \textcolor{red}{0.05} & \textcolor{red}{$4.21\times10^{-4}$} & \textcolor{red}{71 (17)} & \textcolor{red}{3.9} \\
LoTSS-seed20-beta10-t90  & \textcolor{red}{9.18} & \textcolor{red}{$6.77\times10^{58}$} & \textcolor{red}{59.50} & \textcolor{red}{0.14} & \textcolor{red}{$3.23\times10^{-4}$} & \textcolor{red}{123 (2)} & \textcolor{red}{4.24} \\
\hline
\end{tabular}
\end{table*}

\subsection{Plasma beta.} The plasma beta parameter $\beta = \frac{P_{\rm gas}}{P_{\rm mag}}$ (i.e., the magnetization of the non-thermal bubble plasma) affects the emissivity and cooling of the ring but not the bubble population. As our model is scale free as long as the magnetic field is not dynamically dominant \citep[$\beta\geq 1$,][]{wang_odd_2026}, we can vary $\beta$ to investigate its effect on the resulting ORC distribution. The $\beta=1$, 10 and 20 runs share the same \textcolor{red}{2,327,873} surviving bubbles, so any difference between them comes from the radio emission alone. $N_{\rm ORC}$ decreases with increasing $\beta$ in both surveys. For EMU it drops from \textcolor{red}{5.03} at $\beta=1$ to \textcolor{red}{2.7} at $\beta=10$ and \textcolor{red}{1.57} at $\beta=20$, and for LoTSS at 144 MHz from \textcolor{red}{371.75} to \textcolor{red}{268} and \textcolor{red}{170.96}. Over the full range, $N_{\rm ORC}$ therefore changes by a factor of about three for EMU and two for LoTSS. The rings with 1.4 GHz counterparts drop from \textcolor{red}{10.33} to \textcolor{red}{7.46} and \textcolor{red}{6.88}. The median Mach number of the detectable rings rises from \textcolor{red}{4.43} to \textcolor{red}{5.57} and \textcolor{red}{6.02} for EMU and from \textcolor{red}{4.45} to \textcolor{red}{5.07} and \textcolor{red}{5.50} for LoTSS, since a weaker field has to be compensated by a stronger shock for the ring to stay bright enough.

\subsection{Jet-episode duration.} The mean episode duration $t_{\rm duty}$ is how long the jet injects energy into the bubbles. Shorter episodes produce more bubbles. The number of surviving bubbles is \textcolor{red}{4,217,625} for $t_{\rm duty}=4.5$ Myr, \textcolor{red}{2,327,873} for 45 Myr and \textcolor{red}{1,747,253} for 90 Myr. The final expected number $N_{\rm ORC}$ changes \textcolor{red}{less}. For EMU it is \textcolor{red}{4.66}, \textcolor{red}{2.7} and \textcolor{red}{3.20}, and for LoTSS at 144 MHz \textcolor{red}{351.67}, \textcolor{red}{268} and \textcolor{red}{212.84}. Relative to the fiducial run, the EMU numbers change by factors of \textcolor{red}{1.73} and \textcolor{red}{1.18}, and the LoTSS numbers by \textcolor{red}{1.31} and \textcolor{red}{0.79}, while the number of surviving bubbles changes by \textcolor{red}{1.81} and \textcolor{red}{0.75} and The 1.4 GHz counterparts (\textcolor{red}{20.08}, \textcolor{red}{7.46} and \textcolor{red}{9.18}). Most of the additional bubbles from short episodes are therefore too small to be detected. The detectable bubbles are built from many short episodes or a few long ones (a median of \textcolor{red}{23}, \textcolor{red}{3} and \textcolor{red}{2} episodes for EMU), but their total jet-on time changes much less (median \textcolor{red}{102-126} Myr for EMU, \textcolor{red}{122-153} Myr for LoTSS). The properties of the detectable rings also change only slightly. From $t_{\rm duty}=4.5$ to 90 Myr, the median $E_{\rm bub}$ changes from $\textcolor{red}{9.20\times10^{57}}$ to $\textcolor{red}{7.16\times10^{57}}$ erg and the median $R_{\rm ring}$ \textcolor{red}{changes} from \textcolor{red}{47.50} to \textcolor{red}{42.50} kpc for EMU. For LoTSS, $E_{\rm bub}$ rises from $\textcolor{red}{3.01\times10^{58}}$ to $\textcolor{red}{4.66\times10^{58}}$ erg and $R_{\rm ring}$ from \textcolor{red}{62.10} to \textcolor{red}{65.50} kpc, and the median $z_{\rm obs}$ \textcolor{red}{stays at 0.13}.

\subsection{Monte Carlo realizations.} We repeat the fiducial run with a different random seed (34 instead of 20) to estimate the sampling noise. The number of surviving bubbles changes by \textcolor{red}{0.08\%}, from \textcolor{red}{2,327,873} to \textcolor{red}{2,329,631}. $N_{\rm ORC}$ \textcolor{red}{changes more}: it changes for EMU \textcolor{red}{2.7} to \textcolor{red}{3.71}, for LoTSS at 144 MHz \textcolor{red}{268} to \textcolor{red}{298.18}, and for the 1.4 GHz counterparts \textcolor{red}{7.46} to \textcolor{red}{11.97}. The medians of the observable rings agree closely for EMU ($R_{\rm ring}=\textcolor{red}{49.90}$ and \textcolor{red}{49.50} kpc, $z_{\rm obs}=\textcolor{red}{0.09}$ and \textcolor{red}{0.09}). They differ more for LoTSS at 144 MHz ($R_{\rm ring}=\textcolor{red}{65.00}$ and \textcolor{red}{69.70} kpc, $z_{\rm obs}=\textcolor{red}{0.14}$ and \textcolor{red}{0.14})\textcolor{red}{, and agree for the 1.4 GHz sample ($R_{\rm ring}=44.60$ and 45.60 kpc)}. The seed-to-seed scatter in $N_{\rm ORC}$ is below the factor of varying $\beta$ or the shock Mach number (described below), and the 1.4 GHz numbers should be read as estimates within a factor of about two.

\subsection{Uncertainty estimates.} 
\label{sec:uncertainties}
Our model adopts several relations from the literature that carry their own (measured or estimated) uncertainties. To see how these propagate into $N_{\rm ORC}$, we reran the fiducial EMU and LoTSS models for each input separately, each time moving one input to the two ends of its uncertainty and keeping everything else fixed.

The first input is the slope of the radio-AGN incidence plane, $b_\eta=-0.845\pm0.074$ \citep{oshea_shooting_2025}. Because the coefficients of the plane are correlated, we vary $b_\eta$ and adjust $c_\eta$ so that the incidence is unchanged at the center of the fitted data, $\log_{10}M_\star\simeq11.1$ and $\log_{10}\lambda_{\rm jet}\simeq-2.5$. This changes $c_\eta$ to $\textcolor{red}{-21.52}$ and $\textcolor{red}{-21.89}$ for $b_\eta=\textcolor{red}{-0.77}$ and $\textcolor{red}{-0.92}$. 

The second input is the normalization of the hot gas fraction, $(2.09\pm0.14)\times10^{-6}$ \citep{popesso_hot_2026}, which sets the gas density and pressure at all radii in the halo. 


The third input is the peak of the shock Mach-number distribution. \citet{lee_morphology_2024} find that the peak varies between about 2 and 4.5 from one merger to another, and we run the two ends of this range.

The fourth input is the stellar mass function. \citet{bernardi_stellar_2018} list it for two photometric versions, with and without a dust correction, and our fiducial model uses the midpoint of the two. We rerun the model with each version separately.

The remaining four inputs are the parameters of the conditional stellar mass function of \citet{yang_galaxy_2009}, which set halo mass of each host and whether it is a central or a satellite: the satellite normalization $\phi^\ast_{\rm s}$ and slope $\alpha_{\rm s}$, and the central stellar mass $\log M_{\ast,{\rm c}}$ and scatter $\sigma_{\rm c}$ using the errors of their table 4. 

For each input we take half the logarithmic difference between its two runs,
\begin{equation}
\sigma_i = \frac{1}{2}\left|\ln\frac{N_{{\rm ORC},i}^{(1)}}{N_{{\rm ORC},i}^{(2)}}\right|,
\end{equation}
as its contribution to the uncertainty in $\ln N_{\rm ORC}$. This gives $\sigma_b=\textcolor{red}{0.32}$ for the incidence slope ($N_{\rm ORC}=\textcolor{red}{4.05}$ and \textcolor{red}{2.14}), $\sigma_{f_{\rm gas}}=\textcolor{red}{0.02}$ for the gas fraction (\textcolor{red}{3.34} and \textcolor{red}{3.20}). For the Mach number the two runs give $N_{\rm ORC}=\textcolor{red}{9.54}$ and \textcolor{red}{0.58}. These runs sit at the ends of a range rather than at $\pm1\sigma$, so we treat the range as a uniform distribution, whose standard deviation is its half-width divided by $\sqrt{3}$. This gives $\sigma_{\mathcal{M}}=\frac{1}{2}\ln(\textcolor{red}{9.54}/\textcolor{red}{0.58})/\sqrt{3}=\textcolor{red}{0.81}$. Adding the \textcolor{red}{eight} contributions
\begin{equation}
\sigma_{\ln N}=\textcolor{red}{\left(\sum_i\sigma_i^2\right)^{1/2}}=\textcolor{red}{0.89},
\end{equation}
so the prediction is uncertain by a factor of $e^{\textcolor{red}{0.89}}=\textcolor{red}{2.45}$ in either direction, and
\begin{equation}
N_{\rm ORC}(944 \rm{MHz})=\textcolor{red}{2.7^{+3.9}_{-1.6}}
\end{equation}
for the EMU Pilot. This uncertainty is dominated by the shock Mach-number distribution.

The same process gives LoTSS
\begin{equation}
    N_{\rm ORC}^{\rm LoTSS}(144 \rm{MHz})=\textcolor{red}{268^{+331}_{-148}}
\end{equation}
and
\begin{equation}
    N_{\rm ORC}^{\rm LoTSS}(1.4 \rm{GHz})=\textcolor{red}{7.46^{+9.51}_{-4.18}}
\end{equation}
\end{document}